\documentclass[aps,prd,
 reprint,
superscriptaddress,
nofootinbib,
 amsmath,amssymb,
 aps,
]{revtex4-2}

\usepackage{amsmath}
\usepackage{appendix}
\usepackage{graphicx}
\usepackage{dcolumn}
\usepackage{bm}
\usepackage{xcolor}
\usepackage{subfigure} 
\usepackage{url}
\usepackage{hyperref}
\usepackage{multirow}

\newcommand{\DT}{{\Delta T}}
\newcommand{\DE}{{\Delta E}}
\newcommand{\Cl}{{\bm{C}_\ell}}
\newcommand{\planck}{{\it Planck}}

\newcommand{\fsky}{\ensuremath{f_\text{sky}}}

\begin{document}

\preprint{APS/123-QED}

\title{Forecasting ground-based sensitivity to the Rayleigh scattering of the CMB in the presence of astrophysical foregrounds}

\author{Karia R. Dibert}
\affiliation{Department of Astronomy \& Astrophysics, University of Chicago, 5640 South Ellis Avenue, Chicago, IL, 60637, USA}
\affiliation{Kavli Institue for Cosmological Physics, University of Chicago, 5640 South Ellis Avenue, Chicago, IL, 60637, USA}
\author{Adam J. Anderson}
\affiliation{Department of Astronomy \& Astrophysics, University of Chicago, 5640 South Ellis Avenue, Chicago, IL, 60637, USA}
\affiliation{Kavli Institue for Cosmological Physics, University of Chicago, 5640 South Ellis Avenue, Chicago, IL, 60637, USA}
\affiliation{Fermi National Accelerator Laboratory, MS209, P.O. Box 500, Batavia, IL, 60510, USA}
\author{Amy N. Bender}
\affiliation{High-Energy Physics Division, Argonne National Laboratory, 9700 South Cass Avenue., Lemont, IL, 60439, USA}
\affiliation{Kavli Institue for Cosmological Physics, University of Chicago, 5640 South Ellis Avenue, Chicago, IL, 60637, USA}
\author{Bradford A. Benson}
\affiliation{Department of Astronomy \& Astrophysics, University of Chicago, 5640 South Ellis Avenue, Chicago, IL, 60637, USA}
\affiliation{Kavli Institue for Cosmological Physics, University of Chicago, 5640 South Ellis Avenue, Chicago, IL, 60637, USA}
\affiliation{Fermi National Accelerator Laboratory, MS209, P.O. Box 500, Batavia, IL, 60510, USA}
\author{Federico Bianchini}
\affiliation{Kavli Institute for Particle Astrophysics and Cosmology, Stanford University, 452 Lomita Mall, Stanford, CA, 94305, USA}
\affiliation{SLAC National Accelerator Laboratory, 2575 Sand Hill Road, Menlo Park, CA, 94025, USA }
\author{John E. Carlstrom}
\author{Thomas M. Crawford}%
\affiliation{Department of Astronomy \& Astrophysics, University of Chicago, 5640 South Ellis Avenue, Chicago, IL, 60637, USA}
\affiliation{Kavli Institue for Cosmological Physics, University of Chicago, 5640 South Ellis Avenue, Chicago, IL, 60637, USA}
\author{Riccardo Gualtieri}
\affiliation{High-Energy Physics Division, Argonne National Laboratory, 9700 South Cass Avenue., Lemont, IL, 60439, USA}
\author{Yuuki Omori}
\affiliation{Department of Astronomy \& Astrophysics, University of Chicago, 5640 South Ellis Avenue, Chicago, IL, 60637, USA}
\affiliation{Kavli Institue for Cosmological Physics, University of Chicago, 5640 South Ellis Avenue, Chicago, IL, 60637, USA}
\author{Zhaodi Pan}
\affiliation{High-Energy Physics Division, Argonne National Laboratory, 9700 South Cass Avenue., Lemont, IL, 60439, USA}
\author{Srinivasan Raghunathan}
\affiliation{Center for AstroPhysical Surveys, National Center for Supercomputing Applications, Urbana, IL, 61801, USA}
\author{Christian L. Reichardt}
\affiliation{School of Physics, University of Melbourne, Parkville, VIC 3010, Australia}
\author{W. L. Kimmy Wu}
\affiliation{Kavli Institute for Particle Astrophysics and Cosmology, Stanford University, 452 Lomita Mall, Stanford, CA, 94305, USA}
\affiliation{SLAC National Accelerator Laboratory, 2575 Sand Hill Road, Menlo Park, CA, 94025, USA }

\date{\today}

\begin{abstract}
The Rayleigh scattering of cosmic microwave background (CMB) photons off the neutral hydrogen produced during recombination effectively creates an additional scattering surface after recombination that encodes new cosmological information, including the expansion and ionization history of the universe. A first detection of Rayleigh scattering is a tantalizing target for next-generation CMB experiments. We have developed a Rayleigh scattering forecasting pipeline that includes instrumental effects, atmospheric noise, and astrophysical foregrounds (e.g., Galactic dust, cosmic infrared background, or CIB, and the thermal Sunyaev-Zel’dovich effect). We forecast the Rayleigh scattering detection significance for several upcoming ground-based experiments, including SPT-3G+, Simons Observatory, CCAT-prime, and CMB-S4, and examine the limitations from atmospheric and astrophysical foregrounds as well as potential mitigation strategies. When combined with Planck data, we estimate that the ground-based experiments will detect Rayleigh scattering with a significance between 1.6 and 3.7, primarily limited by atmospheric noise and the CIB.
\end{abstract}

\maketitle

    \section{Introduction}


    Cosmic microwave background (CMB) measurements continue to produce ever-tightening constraints on $\Lambda$CDM cosmological parameters. With several next-generation CMB experiments such as SPT-3G+\cite{dibert21}, Simons Observatory \cite{simons19}, and CCAT-prime \cite{ccat21, choi20} deploying soon, and with CMB-S4 \cite{cmbs422} on the horizon, we expect measurements of the primary CMB temperature and polarization power spectra to approach the cosmic variance limit in the coming decades. 
    Further reduction in the uncertainties of cosmological parameters will thus require new and improved measurements of secondary CMB anisotropies.
    Secondary anisotropies are distortions to the primary CMB generated through interactions between the CMB and its environment over the course of its journey from last-scattering to detection. Gravitational lensing of the CMB is one example of a secondary anisotropy.
    
    Secondary CMB anisotropy can also be generated through the interaction of the CMB with neutral hydrogen atoms just after recombination. The usual picture after recombination is of a completely transparent post-recombination universe, but this is not strictly accurate. CMB photons are able to interact with neutral hydrogen atoms through a process known as Rayleigh scattering, in which CMB photons scatter off the induced dipoles of the hydrogen atoms. This interaction has a frequency-dependent cross section which is proportional to $\nu^4$ \cite{yu01, lewis13, alipour15}. Rayleigh scattering can be thought of as a screen just in front of the primary last-scattering surface, providing a frequency-dependent contribution to the primary CMB temperature and polarization signals.
    
    The Rayleigh scattering of the CMB has a number of measurable effects on the CMB temperature and polarization power spectra. On small scales, the increased photon diffusion resulting from Rayleigh scattering leads to the suppression of both temperature and polarization anisotropies. The frequency dependence of the Rayleigh scattering cross section causes the size of the sound horizon to also be frequency dependent, leading to a shift in the locations of acoustic peaks in both the temperature and polarization power spectra. Additionally, Rayleigh scattering boosts E-mode polarization anisotropies on large scales. This results from the shift in the visibility function induced by the scattering of photons after recombination. Effectively, last scattering appears to happen later, at a time when the local temperature quadrupole is larger. This leads to increased E-mode anisotropies on the largest scales \cite{alipour15, lewis13}.
    
    High-sensitivity measurements of Rayleigh scattering have the potential to improve cosmological parameter constraints. It has previously been shown that the cosmological information available from Rayleigh scattering could significantly improve upon the constraint on the primordial helium abundance \cite{alipour15} and on primordial non-Gaussianity constraints \cite{coulton21}. It has also been shown that constraints could be placed on the expansion history and sound speed of the universe at recombination, which could provide information about the parameters upon which these observables depend \cite{beringue21, alipour15}.
    
   A detection of the Rayleigh scattering contribution to the CMB anisotropy is a primary science goal for the next camera on the South Pole Telescope (SPT), called SPT-3G+ \cite{dibert21}. This new higher-frequency camera will observe beyond the peak of the CMB blackbody spectrum, complementing the lower-frequency SPT-3G data \cite{bender18} (see Table \ref{tab:instruments}). To estimate the signal-to-noise achievable on the Rayleigh scattering signal by the combined survey, we require a forecasting pipeline that includes the effects of all potential contaminants. Previous work has forecasted the achievable Rayleigh scattering signal-to-noise of ground-based CMB experiments in the presence of atmospheric emission \cite{lewis13, alipour15, beringue21}. However, the effect of astrophysical foregrounds on Rayleigh scattering sensitivity has only recently begun to be investigated \cite{ccat21}. 
   In this paper, we describe our Rayleigh scattering forecasting pipeline, which includes astrophysical foregrounds in addition to more standard instrumental and atmospheric effects, and estimate the detection significance for upcoming ground-based CMB experiments.

    \section{The Rayleigh Scattering Signal}
    As described in \cite{lewis13}, \cite{lee05} the Rayleigh scattering cross section of photons off ground-state neutral hydrogen is given by a frequency-dependent modification to the Thomson scattering cross section:
    
    \begin{multline}
       \sigma_R \approx \sigma_T \Bigg [ \left ( \frac{\nu}{\nu_\text{eff}} \right )^4  + \frac{648}{243}\left ( \frac{\nu}{\nu_\text{eff}} \right )^6 \\ + \frac{1299667}{236196}\left ( \frac{\nu}{\nu_\text{eff}} \right )^8 +... \Bigg ], 
    \end{multline}
    where $\sigma_T$ is the Thomson cross section and $\nu_\text{eff}$ is roughly the frequency of an H ionizing photon. The initial $\nu^4$ term largely dominates, and will be the only Rayleigh scattering cross section considered in this analysis. This is because $\nu \ll \nu_{\text{eff}}$ for any millimeter or submillimeter frequency.
    
\begin{figure}
        \centering
        \includegraphics[width=\columnwidth]{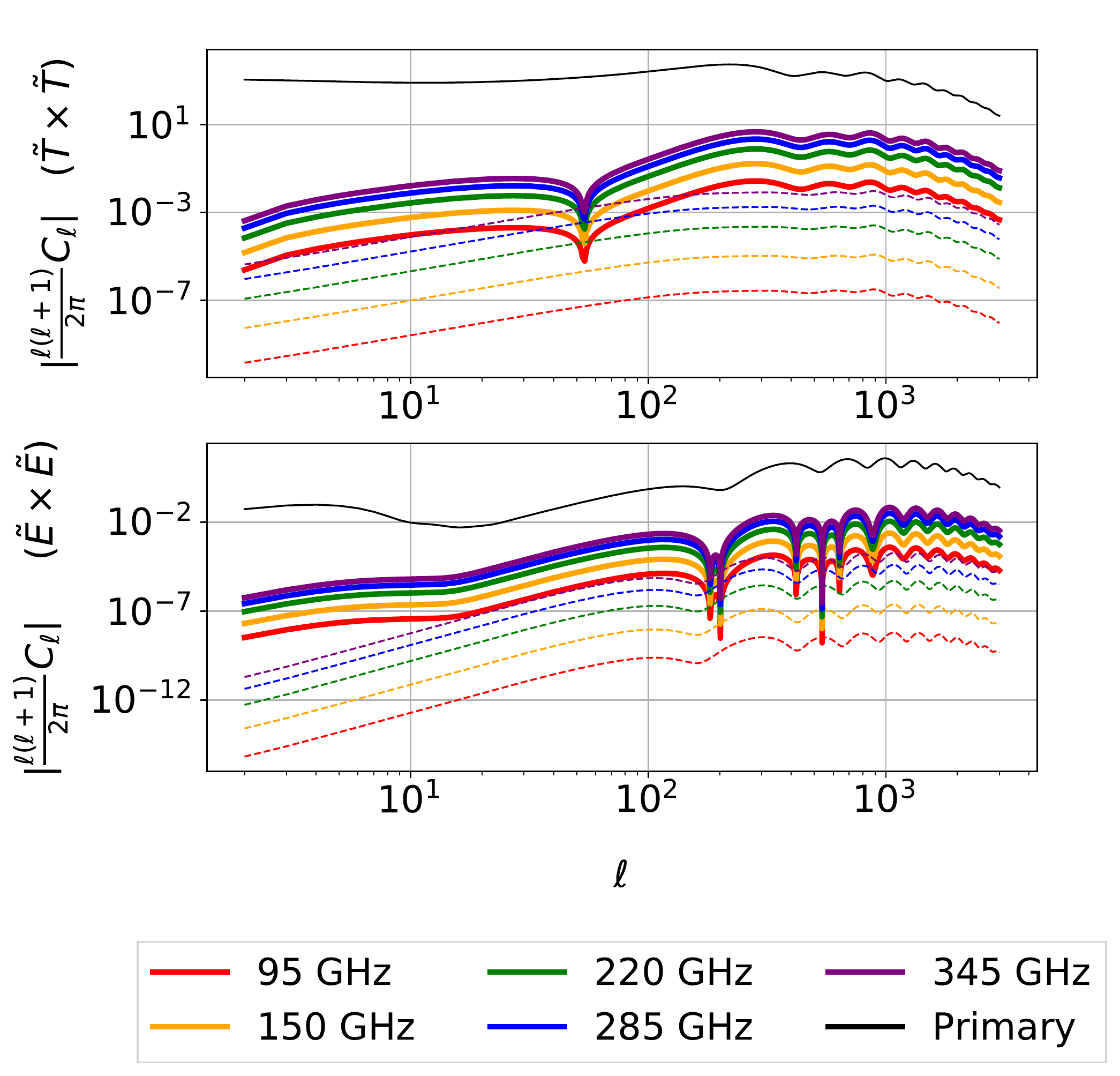}
        \caption{CAMB predictions of Rayleigh scattering power spectral contributions for SPT-3G and proposed SPT-3G+ bands.  \textit{Top}: Rayleigh scattering contributions to the CMB temperature power spectrum. The absolute value of the primary-primary temperature power spectrum is shown in black, while solid-colored lines represent the absolute value of the primary-Rayleigh cross-spectrum for each frequency band. Dotted lines represent the absolute value of the Rayleigh auto-spectrum for each frequency band. \textit{Bottom:} Rayleigh scattering contributions to the CMB E-mode polarization power spectrum. Black, solid-colored, and dotted lines have meanings analogous to those of the corresponding lines in the top panel.}
        \label{fig:camb}
    \end{figure}

    We model the total CMB temperature signal as a sum of a primary CMB component and a frequency-dependent distortion induced by Rayleigh scattering: $\widetilde{T} = T + \Delta T$.  Here, $\widetilde{T}$ represents the total Rayleigh-distorted temperature signal, $T$ represents the primary CMB temperature signal without Rayleigh distortion, and $\DT$ represents the frequency-dependent Rayleigh scattering contribution to the temperature signal. This means that the total CMB temperature power spectrum of a Rayleigh scattered CMB has the form:
    
    \begin{equation}
    \begin{split}
    & C_\ell^{\widetilde{T} \widetilde{T}} =  \langle \widetilde{T} \, \, \widetilde{T} \rangle\\
    & =  \langle T \, \, T \rangle + 2 \langle T \, \, \DT \rangle + \langle \DT \, \, \DT \rangle.
    \end{split}
    \end{equation}
A similar form can also be written for the Rayleigh scattering distortion of the E-mode polarization power spectrum. Using the modified version of CAMB described in \cite{lewis13} to model Rayleigh scattering power spectra, we calculate the Rayleigh cross- and auto-spectra expected for the SPT-3G and SPT-3G+ observing bands, shown in Figure \ref{fig:camb}. The solid black lines indicate the absolute values of the primary CMB temperature and E-mode polarization power spectra $C_\ell^{TT}$ and $C_\ell^{EE}$ respectively. The solid colored lines indicate the absolute values of the primary-Rayleigh temperature and polarization cross-spectra $C_\ell^{T\DT}$ and $C_\ell^{E \DE}$ respectively. Note the $\nu^4$ dependence of the amplitudes of these cross-spectra. Dotted colored lines indicate the absolute values of the Rayleigh temperature and polarization auto-spectra $C_\ell^{\DT \DT}$ and $C_\ell^{\DE \DE}$ respectively. The auto-spectra have a $\nu^8$ dependence, and an amplitude so much lower than their cross-spectrum counterparts as to be essentially negligible in comparison. Indeed, in the following section we negelect Rayleigh auto-spectrum terms throughout our derivation of the total Rayleigh scattering signal-to-noise. This assumption will turn out to be well-motivated, as the Rayleigh auto-spectrum amplitude is several orders of magnitude less than the already difficult-to-detect Rayleigh cross-spectrum.
   
    \section{Methods} \label{methods}
    
    Our method of computing the total Rayleigh signal-to-noise at each multipole consists of two steps. The first is to separate the Rayleigh scattering signal from the primary CMB signal in the presence of noise and foregrounds. This component separation results in expected signal and noise power spectra for each primary CMB auto-spectrum, primary-Rayleigh cross-spectrum, and Rayleigh auto-spectrum. The second step uses these values to compute the total Rayleigh scattering signal-to-noise via the Fisher formalism.
    
    \subsection{Component separation}
    We employ a constrained linear combination algorithm similar to the one described in \cite{remazeilles18} to separate the Rayleigh scattering signal from the primary CMB signal. For a set of maps at various frequencies, this method identifies linear combinations of maps with the minimum possible variance, one of which 1) is an unbiased representation of the Rayleigh scattering signal, and 2) contains formally zero primary CMB signal, and the other of which is an unbiased representation of primary CMB with no response to Rayleigh signal. For a set of temperature and E-mode maps at frequencies $\nu$, $\bm{X} \equiv [T_\nu,E_\nu]$, the best estimate for orthogonal primary CMB and Rayleigh maps $\hat{\bm{Y}} \equiv [\hat{T},\hat{E},\Delta \hat{T},\Delta \hat{E}]$ are given by:
\begin{equation}
\bm{Y} = \bm{w}^t \bm{X},
\end{equation}
where
\begin{equation}
\bm{w}^t= (\bm{a}^t (\bm{C}+\bm{N})^{-1} \bm{a})^{-1} \ \bm{a}^t (\bm{C}+\bm{N})^{-1},
\label{eqn:wts}
\end{equation}
$\bm{a}$ is a 2-by-\# of bands matrix representing the frequency dependence of the primary CMB and Rayleigh signals, and $\bm{C}$ and $\bm{N}$ are the band-band signal and noise covariance matrices. If we choose to work in multipole space, and we assume all sources of signal and noise are statistically isotropic and Gaussian-distributed, then we can assume $\bm{C}$ and $\bm{N}$ are only functions of $\ell$ (not $m$) and are diagonal in $\ell$ space. In this case, we can perform the calculation independently at each value of $\ell$ and write $\bm{C}$ as $\bm{C}_\ell (\nu_i, \nu_j)$, and similarly with $\bm{N}$.
        
    The signal covariance matrix $\bm{C}_\ell (\nu_i, \nu_j)$ is constructed from the CAMB-modeled Rayleigh and primary CMB power spectra in the previous section. 
    This means, for example:
    
    \begin{multline}
    \Cl^{\widetilde{T}\widetilde{E}}(\nu_i, \nu_j) = C_\ell^{TE}(\nu_i,\nu_j) + C_\ell^{T\DE}(\nu_i,\nu_j) \\ 
    + C_\ell^{\DT E}(\nu_i,\nu_j) + C_\ell^{\DT \DE}(\nu_i,\nu_j).
    \end{multline}

    For the purposes of this analysis, we assume the final auto-spectrum term to be negligible, meaning that the each matrix entry is a sum of a frequency-independent primary CMB term and two $\nu^4$-dependent Rayleigh-primary cross-spectrum terms.  The noise part of the covariance matrix, $\bm{N}_\ell(\nu_i, \nu_j),$ is constructed using models for detector noise, atmospheric emission, and galactic/extragalactic foregrounds. These models are discussed extensively in Section~\ref{noise}.

    \subsection{Fisher calculation}

    We compute the total Rayleigh signal-to-noise using the Fisher formalism. This method produces a combined signal-to-noise value that takes into account correlations between the various primary-Rayleigh cross-spectra. Using the outputs of the component-separation procedure in the previous section, $\hat{\bm{Y}} \equiv [\hat{T},\hat{E},\Delta \hat{T},\Delta \hat{E}]$, we construct our best estimates of the Rayleigh-primary cross-spectra, for example:
    \begin{eqnarray}
        \hat{C}_\ell^{T\DT} &=& \frac{1}{2 \ell + 1} \sum_{m=-\ell}^\ell \hat{T}_{\ell m} \Delta \hat{T}_{\ell m} \\
        \nonumber &=&  \frac{1}{2 \ell + 1} \sum_{m=-\ell}^\ell \bm{w}^t_{T,\ell} \bm{X}_\ell \bm{X}_\ell \bm{w}_{\Delta T,\ell},
    \end{eqnarray}
    where $\bm{w}^t_{T,\ell}$ and $\bm{w}^t_{\Delta T,\ell}$ are the $T$ and $\Delta T$ components of the weights defined in Equation~\ref{eqn:wts}. We note that the expectation value of this estimate is equal to
    \begin{equation}
        \langle \hat{C}_\ell^{T\DT} \rangle = C_\ell^{T\DT} + N_\ell^{T\DT},
    \end{equation}
    where $N_\ell^{T\DT} = \bm{w}_{T,\ell}^t \bm{N}_\ell^{\widetilde{T}\widetilde{T}}\bm{w}_{\Delta T,\ell}$. We also note that the $T \Delta E$ and $\Delta T E$ versions of this have no noise bias term. We thus adopt as our data vector:
    \begin{equation}
    \bm{d}_\ell = \left [ 
    \hat{C}_\ell^{T\DT} - N_\ell^{T\DT}, 
    \hat{C}_\ell^{T\DE},  \hat{C}_\ell^{\DT E}, \hat{C}_\ell^{E\DE} - N_\ell^{E\DE} \right ].
    \end{equation}
   Our model of this data vector is that it is
   equal to some constant amplitude $A$ times the model cross-spectra $\bm{s_\ell}$ calculated by CAMB plus noise:
    
    \begin{equation}
    \label{S}
    \begin{split}
    \bm{d}_\ell & = A\bm{s}_\ell + \bm{n}_\ell, \\
    \bm{s}_\ell & = \left [ C_\ell^{T\DT}, C_\ell^{T\DE},  C_\ell^{\DT E}, C_\ell^{E\DE}  \right ].
    \end{split}
    \end{equation}
    The total Rayleigh scattering cross-spectrum signal-to-noise is then given by the signal-to-noise on the parameter $A$. The Fisher matrix, which in this one-parameter case is a single value $F_\ell$, is defined:
    
    \begin{equation}
    \label{eqn:fisher}
    F_{\ell} = -\frac{\partial^2 \ln{\mathcal{L}}}{\partial A^2} ,
    \end{equation}
    where $\mathcal{L}$ is the likelihood function:
    
    \begin{equation}
    \label{eqn:likelihood}
    \mathcal{L_\ell} \propto \exp \left[ -\frac{1}{2} [\bm{d}_\ell - A\bm{s}_\ell]^\mathbf{\intercal} \mathbf{\Xi}^{-1}_\ell [\bm{d}_\ell-A\bm{s}_\ell] \right].
    \end{equation}
    Here $\mathbf{\Xi}_\ell$ is the covariance matrix of the primary-Rayleigh cross-spectra, whose elements are:
    
     \begin{equation}
     \label{eqn:xi}
    \begin{split}
    \mathbf{\Xi}_{\ell, (AB, CD)} = & \frac{1}{(2\ell+1)f_\text{sky}} \left[ \left ( C_{\ell}^{AC} +N_{\ell}^{AC} \right) \left ( C_{\ell}^{BD} + N_{\ell}^{BD} \right ) \right. \\+ 
    & \left. \left ( C_{\ell}^{AD} + N_{\ell}^{AD} \right ) \left ( C_{\ell}^{BC} + N_{\ell}^{BC} \right ) \right],
    \end{split}
    \end{equation}
    with $ A,B,C,D \in \{T, \DT, E, \DE \}$. 
    For example, the Rayleigh temperature cross-spectrum on-diagonal term $\bm{\Xi}_{\ell,(T\DT, T\DT)}$ is:
    
    \begin{equation}
     \frac{ (C_\ell^{TT}+N_\ell^{TT}) (C_\ell^{\DT \DT} + N_\ell^{\DT \DT}) + (C_\ell^{T\DT} + N_\ell^{T\DT})^2}{(2 \ell + 1) f_{\text{sky}}} .
    \end{equation}
    Note that this is equivalent in form to the expression for the temperature cross-spectrum Fisher noise given in Equation 18 of \cite{remazeilles18}. Inserting Equations \ref{eqn:likelihood} and \ref{eqn:xi} into Equation \ref{eqn:fisher}, the single Fisher matrix element reduces to:
    
    \begin{equation}
    F_\ell = \bm{s^{\intercal}} \bm{\Xi}^{-1} \bm{s}
    \end{equation}
    The signal-to-noise on $A$ at a given $\ell$ is then:
    
     \begin{equation}
     S/N(\ell) = \sqrt{F_\ell} = \sqrt{ \bm{s}_\ell^\intercal \bm{\Xi}^{-1}_\ell \bm{s}_\ell}.
    \end{equation}
    We assume noise and foregrounds to be Gaussian and hence uncorrelated between multipoles, however we note that some foregrounds are likely to be mildly non-Gaussian. Therefore the signal-to-noise forecasts presented below should be taken as upper bounds. The cumulative Rayleigh signal-to-noise over all multipoles is the quadrature sum of the signal-to-noise at each multipole:
    
    \begin{equation}
    S/N = \left [ \sum_{\ell>50}  \bm{s}_\ell^\intercal \bm{\Xi}^{-1}_\ell \bm{s}_\ell \right ]^{1/2}. 
    \label{eqn:snr}
    \end{equation}
    
    In Equation \ref{eqn:snr}, we impose a minimum multipole on the sum.
    Beyond the limitations from atmospheric noise and large-angular-scale galactic foregrounds (which are accounted for in the Fisher forecast), the minimum multipole accessible by a ground-based experiment is also limited by the partial sky coverage and potentially by contamination from terrestrial features picked up by the far sidelobes of the beam. We choose $\ell_\text{min} = 50$, which is well above the fundamental limit set by the size of the $\fsky = 0.03$ patch that is the main survey field for SPT-3G and the planned main survey field for SPT-3G+. The difference in total $S/N$ between $\ell_\text{min} = 50$ and no minimum is negligible ($< 1$\%). Equation~\ref{eqn:snr} with $\ell_\text{min} = 50$ is what we report as the ``total Rayleigh signal" for a given experiment and set of foregrounds.

\section{Noise Model} \label{noise}

Our noise model includes contributions from instrumental detector noise, atmospheric emission, and galactic and extragalactic foregrounds. Our models for detector noise and atmospheric emission are similar to those presented in \cite{lewis13}, \cite{alipour15}, and \cite{beringue21}. 
Each foreground is modeled as an independent noise component with its own covariance matrix. The noise input to the component separation algorithm described above is the sum of these foreground covariance matrices, the atmospheric covariance matrix, and the diagonal matrix representing the detector noise. Foregrounds are broadly grouped into galactic and extraglactic sources. The following subsections will describe the functional forms of all noise components considered in our analysis, while the foreground model parameters are included in the appendix to this paper. 
\subsection{Instrument detector noise}
\label{sec_white_noise}

 For an instrument observing at a set of frequencies $\nu_i$, with the contribution to map noise from detectors in each band equal to $N_{\text{det}_i}$, the detector noise covariance matrix is just the diagonal matrix:
\begin{equation}
\bm{N}^\text{det}(\nu_i,\nu_j) =  N^\text{det}_i \delta_{ij}
\end{equation}
Table \ref{tab:instruments} gives the estimated full-survey detector noise values ($N_\text{det}$) for SPT-3G/SPT-3G+ along with other upcoming CMB experiments.

\subsection{Atmospheric emission}
\label{sec:atmo}
 All ground-based CMB experiments must consider emission from atmospheric water vapor as a major source of signal contamination. Similarly to \cite{beringue21}, we define for each frequency band and observing site a characteristic $\ell_\text{knee}$ below which white detector noise is overtaken by noise from atmospheric water vapor, which we model as a power law in $\ell$ with index $\alpha$. Atmospheric noise in a given frequency band is modeled as:

\begin{equation}
\bm{N}^\text{atmos}_\ell (\nu_i) =  N^\text{det}_i \Big( \frac{\ell_\text{knee}(\nu_i)}{\ell} \Big)^{\alpha}.
\end{equation}

With this in mind, the covariance matrix for atmospheric noise is:

\begin{equation}
\bm{N}^\text{atmos}_\ell (\nu_i,\nu_j) = N^\text{det}_i \Big( \frac{\ell_\text{knee}(\nu_i)  }{\ell} \Big)^{\alpha} \delta_{ij}.
\label{eqn:atmocov}
\end{equation}
Atmospheric noise parameters for SPT along with several upcoming ground-based CMB experiments are given in Table \ref{tab:instruments}. Note that by this definition, we assume that the atmospheric noise is totally uncorrelated between bands; we explore the effect of the opposite assumption (100\% correlation between bands) in Section~\ref{sec:atmocorr}.

\begin{figure}[!ht]
    \centering
     \includegraphics[width=\columnwidth, trim=0 0 0 65, clip]{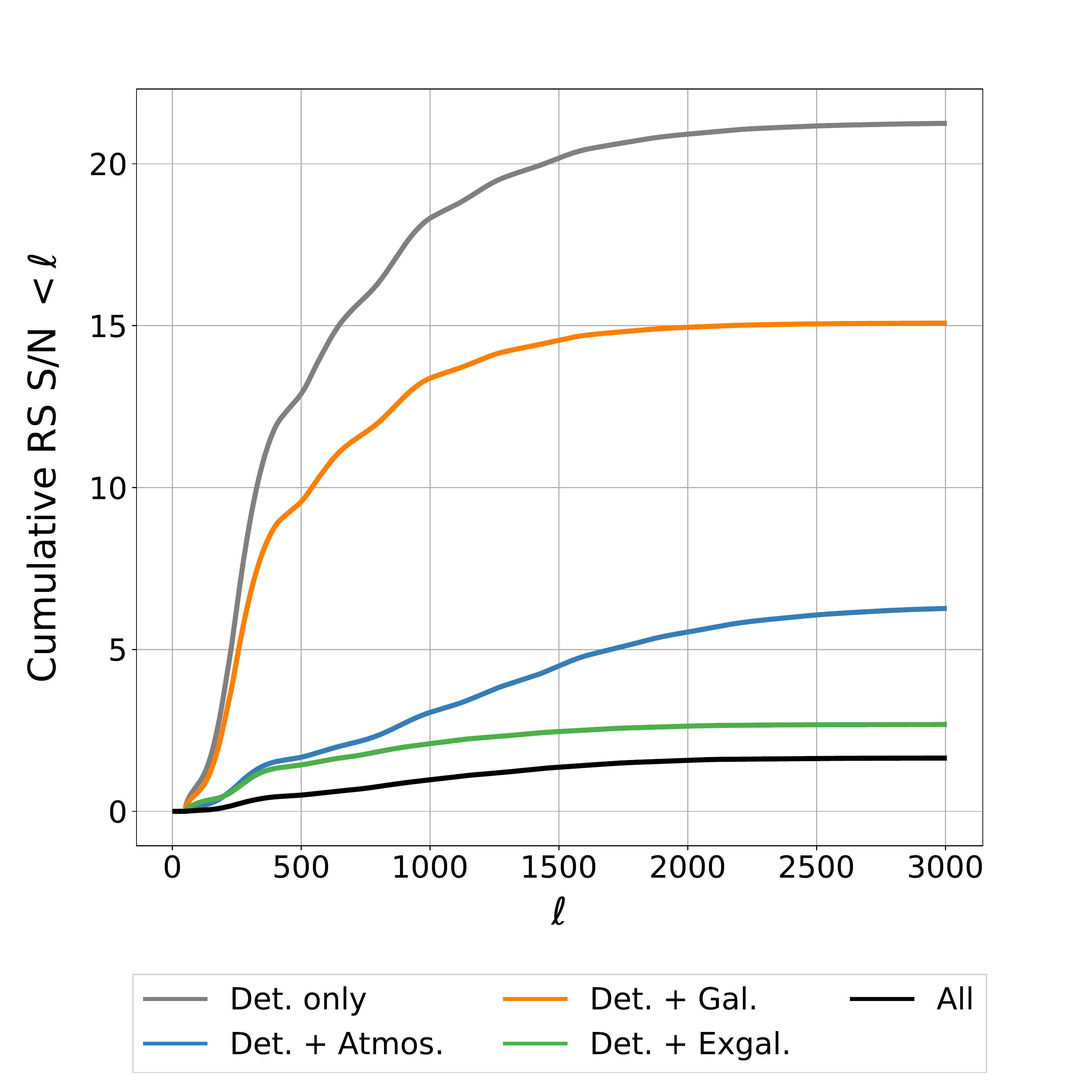}
    \caption{Effect of each noise component on cumulative Rayleigh scattering signal-to-noise for SPT-3G and SPT-3G+ data combined with Planck. Atmosphere and extragalactic foregrounds strongly limit the achievable signal-to-noise at low $\ell$, while extragalactic foregrounds alone become the dominant limiting factor as $\ell$ increases.}
    \label{fig:spt4_breakdown}
\end{figure}

\begin{figure*}
\begin{centering}
\subfigure[Detector noise only]{\label{fig:spec_det} 
 \includegraphics[width=0.95\columnwidth, trim=0 0 0 65, clip]{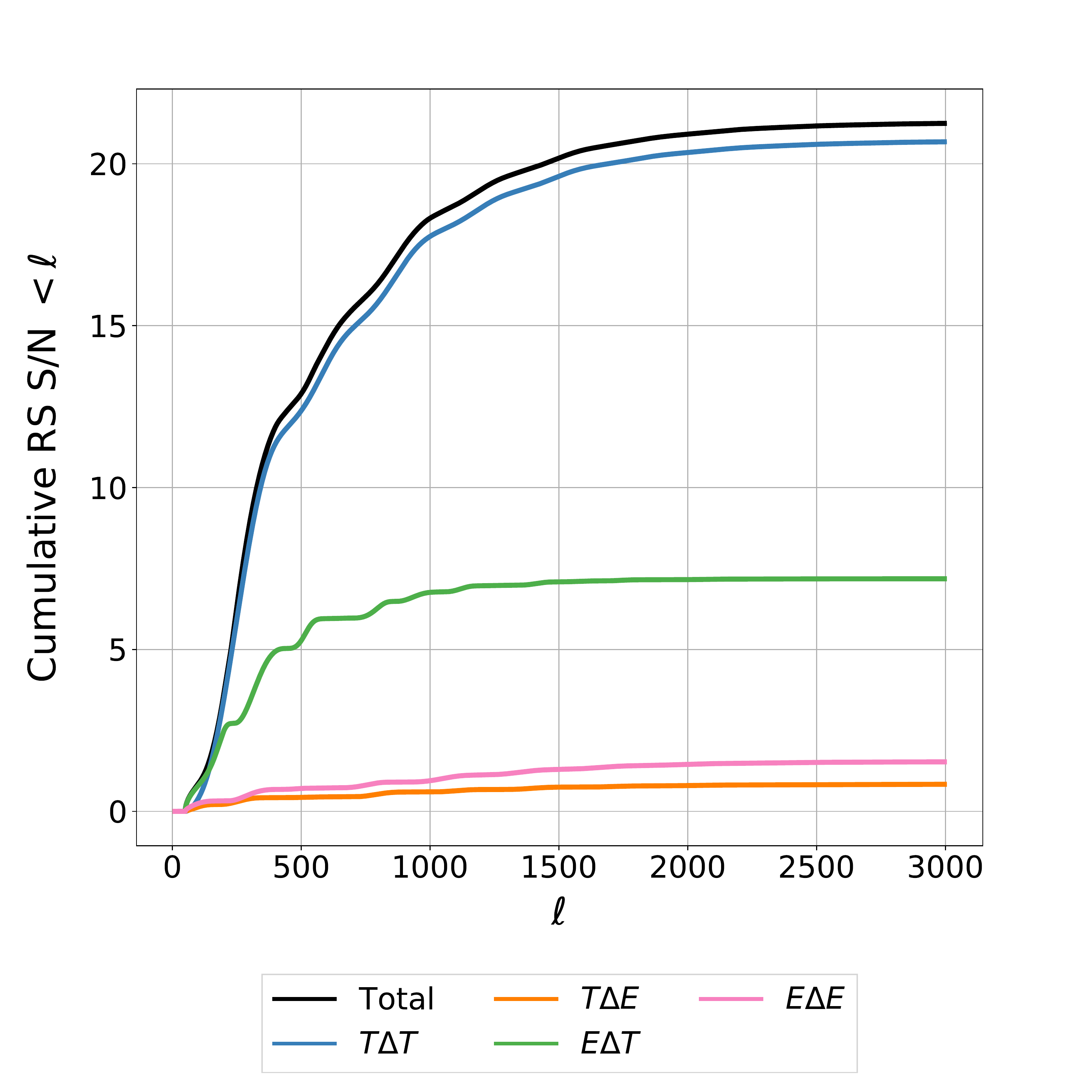}}
\subfigure[Detector noise, atmosphere, and all foregrounds]{\label{fig:spec_all} 
 \includegraphics[width=0.95\columnwidth, trim=0 0 0 65, clip]{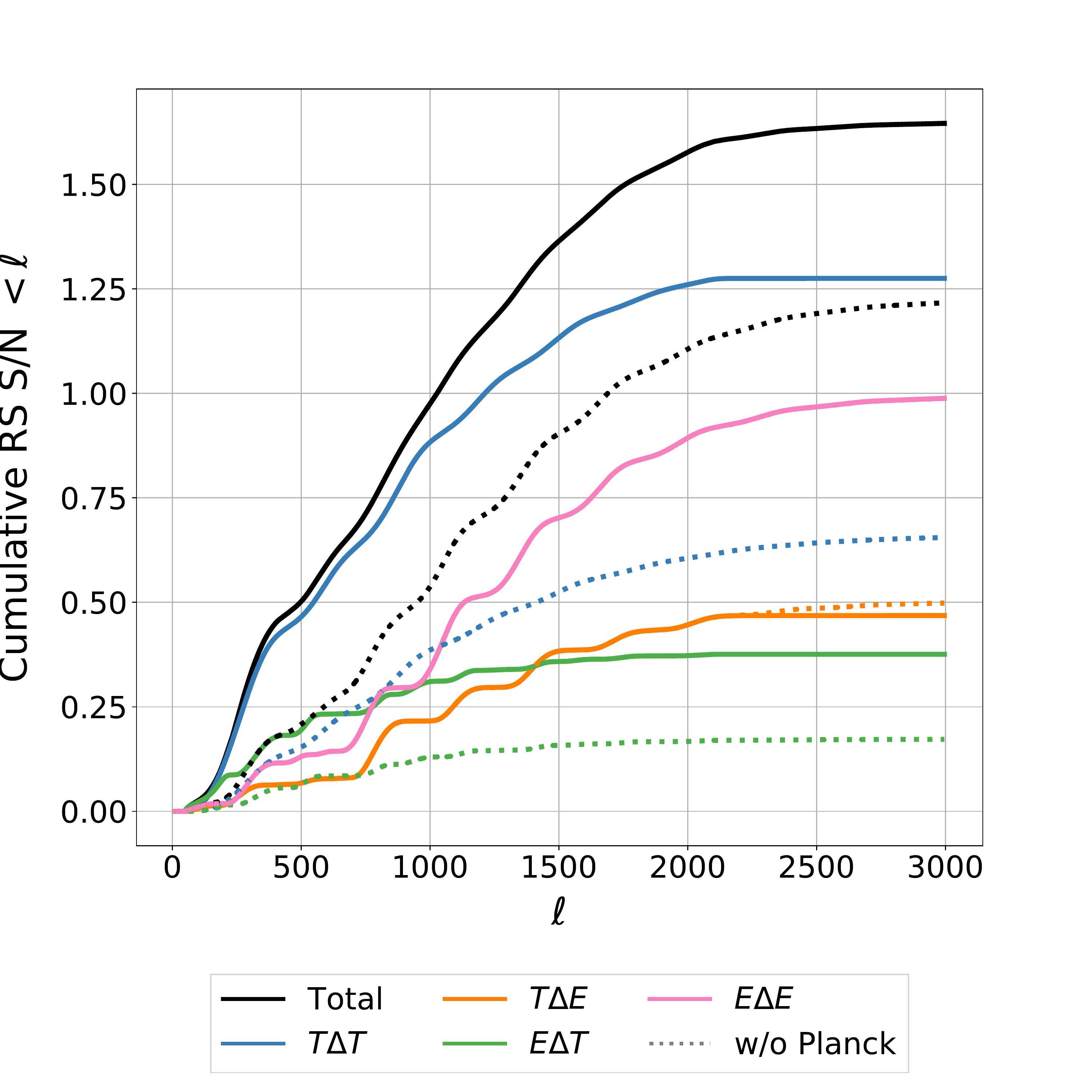}}
\caption{
\label{fig:spt4_spectra} 
Rayleigh signal-to-noise for the combination of SPT-3G, SPT-3G+, and Planck broken down by spectrum, \textit{Left} in the absence of foregrounds and \textit{Right} including all foregrounds and atmosphere. These spectra are correlated, which causes the total combined Rayleigh signal-to-noise for SPT to be less than the quadrature sum of the signal-to-noise of the individual spectra. As expected, the total cumulative Rayleigh signal-to-noise is dominated by that of the $T\DT$ cross-spectrum. The $T\DT$ and $E\DT$ signal-to-noise are severely diminished by the addition of foregrounds, most notably extragalactic foregrounds. The $T\DE$ and $E\DE$ signal-to-noise are less affected, and this slight degradation is mostly due to galactic dust. The dotted lines in the right panel show the signal-to-noise for each spectrum when Planck data is excluded. These lines are not included in the left panel because they are visually indistinguishable from the corresponding solid lines.
}
\end{centering}
\end{figure*}

\subsection{Galactic sources}
\label{sec_gal_sources}

Emission from dust grains in our Galaxy is a known contaminant to measurements of the CMB.
The contribution of galactic dust emission to the $TT$ or $EE$ spectra is modeled by a power law in $D_\ell \equiv \frac{\ell (\ell + 1)}{2 \pi} C_\ell$:
\begin{equation}
D_\ell (\nu) = A_\text{dust}(\nu)\Big( \frac{\ell}{80} \Big)^{\alpha}.
\label{eqn:gal}
\end{equation}
Following \cite{raghunathan22}, we use the publicly available Python Sky Model (pySM) simulations \citep{thorne17, zonca21} to estimate $A_\text{dust}$ and $\alpha$ at 145~GHz for $TT$ and $EE$.
Galactic dust temperature and polarization amplitudes for 
all experiments considered here
are given in Table \ref{tab:foregrounds} in the Appendix to this paper.
We assume the $TE$ spectrum for galactic dust to be the geometric mean of the $TT$ and $TE$ factors times a correlation coefficient of 0.35. We scale this amplitude to other frequency bands using a modified blackbody approximation, the details of which are discussed in Appendix A.2. Assuming full correlation of the galactic dust signal between frequency bands, the noise covariance matrix for galactic dust is:
\begin{equation}
\bm{N}^\text{dust}_\ell (\nu_i, \nu_j) = \frac{2 \pi}{\ell(\ell + 1)}\sqrt{D_\ell (\nu_i) D_\ell(\nu_j)},
\end{equation}
where the prefactor converts from $D_\ell$ to $C_\ell$ space.

Galactic synchrotron emission is also generally considered to be an important contaminant for CMB experiments, particularly at frequencies below the peak of the CMB blackbody spectrum. 
We model synchrotron using a power law as in Equation \ref{eqn:gal}, 
with 
temperature and polarization amplitudes for each experiment again given in Table \ref{tab:foregrounds}.
See Appendix A.2 for further discussion of these values and their scaling to other frequency bands.

\subsection{Extragalactic sources}
\label{sec_extragal_sources}

Our extragalactic foreground model consists of thermal Sunyaev-Zel'dovich (tSZ) and cosmic infrared background (CIB) components, as well as extragalactic radio sources. The tSZ component is modeled as a power law in $\ell$:

\begin{equation}
\label{eqn:fg_power}
 D_{\ell}(\nu) = A_\text{tSZ}(\nu) \left( \frac{\ell}{3000}\right) ^{\alpha},
\end{equation}
where $A_{\text{tSZ}} = 4\, \mu \mathrm{K}^2$ and $\alpha = 0$ at 150 GHz, as seen in Table \ref{tab:foregrounds}. The method for scaling the tSZ amplitude to other frequency bands is described in Appendix A.3. We neglect any polarized tSZ component. 

Modeling the CIB is a challenging task that has been the subject of many detailed studies (e.g., \cite{lagache03,shang12}). For this work, we are primarily interested in: 1) correctly reproducing the total power and frequency scaling of the CIB reported in the literature, including frequency decorrelation; and 2) being able to separate the clustered and shot-noise components of the CIB. To this end, we have modeled the CIB as originating from two separate infinitely thin screens at redshifts $z=0.5$ and $z=3.5$. At each redshift, there is a clustered component and a shot-noise (``Poisson") component, for a total of four independent components. The Poisson component is flat in $C_\ell$, while the $\ell$-space shape of the clustered CIB components is assumed to follow a power law like that in Equation \ref{eqn:fg_power}, but with an index $\alpha=-1.2$, following, e.g., \cite{george15}. 
The amplitudes of the four CIB components are given in Table \ref{tab:foregrounds}.
Scaling of these amplitudes to other frequency bands is described in Appendix A.3. While this model is clearly \textit{ad hoc} and unphysical, it does reproduce key results in the literature for clustered and Poisson CIB power at 150 and 220 GHz \cite{george15} and the degree of correlation in CIB power between bands from 95 to 1200 GHz \cite{viero19}. The clustered and Poisson CIB are considered to be unpolarized. While the clustered component is unpolarized by construction, the Poisson component has been suggested to be $4\%$ polarized as an upper bound \cite{gupta19}. We have repeated these forecasts for a case in which the Poisson CIB component is $4\%$ polarized and found negligible change in the results of the forecasts.

Extragalactic radio sources are primarily a contaminant at low frequencies. While their effect on the high-frequency SPT-3G+ bands is negligible, their inclusion is necessary when forecasting the Rayleigh scattering sensitivity of other planned experiments. We assume the clustering power of radio sources to be negligible and only forecast the Poisson signal, adopting a value of $A_\text{radio} = 0.17\, \mu \mathrm{K}^2$ at 150 GHz, as seen in Table \ref{tab:foregrounds}. This is lower than the measured value in, e.g., \cite{george15}, because we assume a flux cut of 1 mJy (roughly the $5 \sigma$ detection threshold in the SPT-3G 150 GHz band), compared to roughly 6 mJy in that work. When we forecast for other experiments, we keep this power constant despite the fact that those experiments will have slightly different source detection thresholds. We have checked that using the Simons Observatory 145 GHz detection threshold of roughly 2.7 mJy (which results in a radio Poisson amplitude of $A_\text{radio} = 0.51\, \mu \mathrm{K}^2$) has no measurable effect on our results.\footnote{We note that the dusty source Poisson amplitude is insensitive to source cut threshold down to below 1 mJy at 150 GHz \citep[e.g.,][]{cai13}, at which point the number of sources masked approaches the number of independent resolution elements in the map---i.e., the dusty source Poisson power is dominated by sources at or below the confusion limit for a $\sim$1-arcmin beam.} We assume extragalactic radio sources to be $3\%$ polarized following \cite{datta18}, \cite{gupta19}. The scaling of this model to other frequency bands is again detailed in Appendix A.3.

Using the above expressions for each $D_\ell$, the covariance matrix for each extragalactic foreground component (assuming 100\% correlation between bands) can be expressed:

\begin{equation}
\bm{N}^\text{fg}_\ell (\nu_i, \nu_j) = \frac{2 \pi}{\ell(\ell + 1)} \sqrt{D_\ell (\nu_i) D_\ell(\nu_j)},
\end{equation}
where the prefactor again converts from $D_\ell$ to $C_\ell$ space.

\section{Results and Discussion}
\label{sec_results_discussion}

Using the component separation and Fisher calculation methods described in Section \ref{methods}, with the noise part of the covariance matrix constructed from the components described in Section \ref{noise}, the total Rayleigh signal-to-noise at each multipole can be calculated. We first present these forecasts for SPT, including the current SPT-3G camera and the planned SPT-3G+ camera. Throughout this section, we assume that all experiments will perform a joint analysis with Planck data, however we will quantify the impact of this assumption on our forecasts. For SPT, which observes approximately 3\% of the sky, we include Planck data from the same sky patch. Quantitatively, this means we add rows and columns to our correlation matrix corresponding to Planck's frequency bands, but maintain $\fsky = 0.03$ throughout the Fisher calculation.

The frequency bands used in this forecast include SPT-3G's 95, 150, and 220 GHz bands, SPT-3G+'s, 225, 285, and 345 GHz bands and Planck's 30, 44, 70, 100, 143, 217, 353, 545, and 857 GHz bands. 
Detector noise values for the SPT bands are given in Appendix A, while Planck detector noise values come from Table 4 of \cite{planck20b}. Being a space-based experiment, Planck has no atmospheric noise component.

Figure \ref{fig:spt4_breakdown} shows the resulting SPT Rayleigh signal-to-noise for four scenarios: (1) detector noise only in gray, (2) detector noise plus atmospheric emission in blue, (3) detector noise plus galactic dust in green, and (4) detector noise plus extragalactic sources in orange. The black line represents the total Rayleigh signal-to-noise when all noise components are considered together. The curves on this plot represent the cumulative Rayleigh signal-to-noise up to each multipole $\ell$. This is the result of Equation 6 for a given $\ell$, and is equal to the quadrature sum of all individual multipole signal-to-noise values up to and including $\ell$.

It is immediately apparent that extragalactic foregrounds have the most dramatic effect on the Rayleigh scattering signal-to-noise for SPT. This effect is comparable to the effect of the atmosphere at low multipoles, but persists through higher multipoles at which atmospheric contamination is of less concern. Figure \ref{fig:spt4_breakdown} demonstrates that for SPT, the Rayleigh scattering detection is limited by extragalactic foregrounds more than it is limited by atmospheric noise or detector noise.

\begin{figure*}
\begin{centering}
\subfigure[Detector noise only]{\label{fig:det} 
 \includegraphics[width=0.95\columnwidth, trim=0 0 0 65, clip]{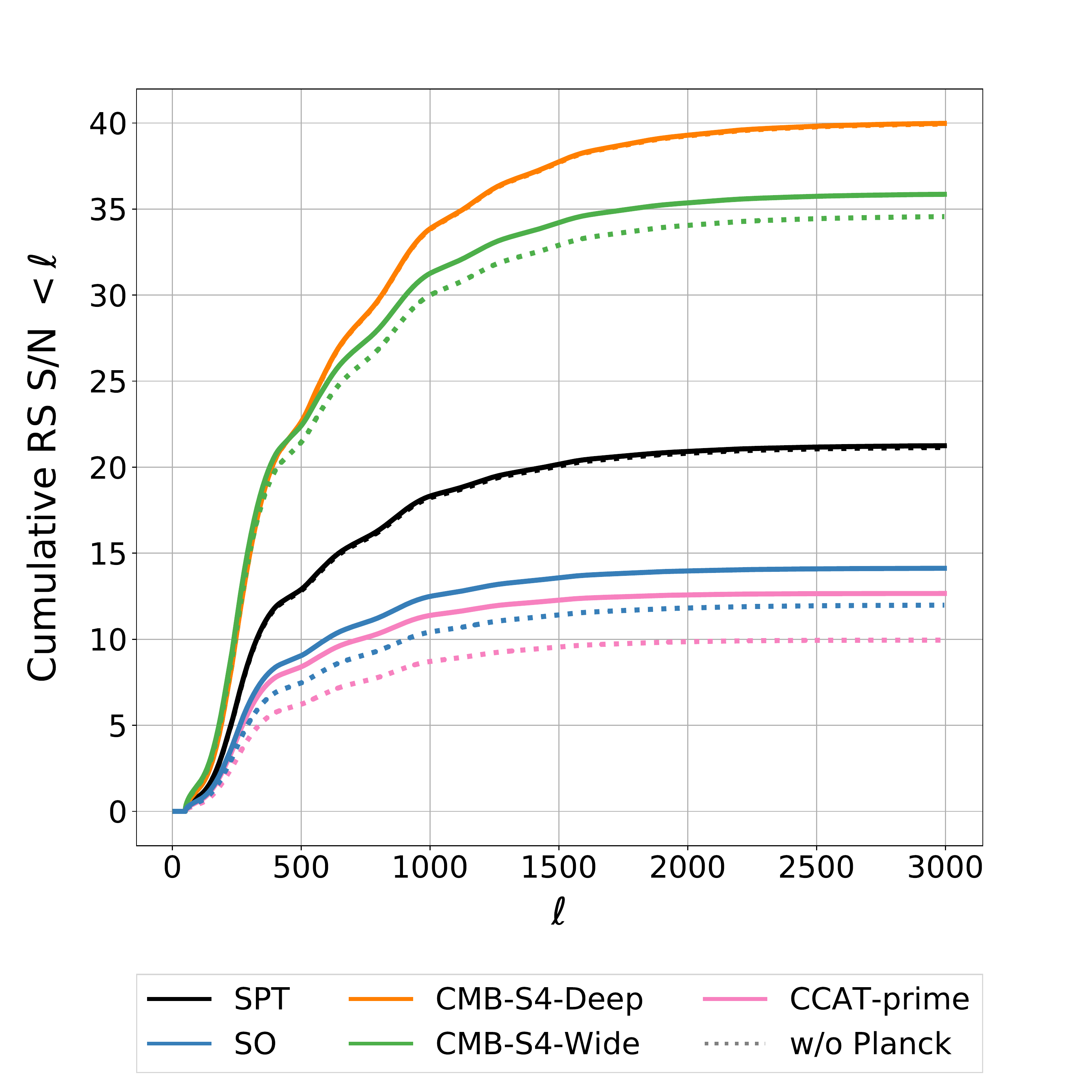}}
\subfigure[Detector noise and atmosphere]{\label{fig:atm} 
 \includegraphics[width=0.95\columnwidth, trim=0 0 0 65, clip]{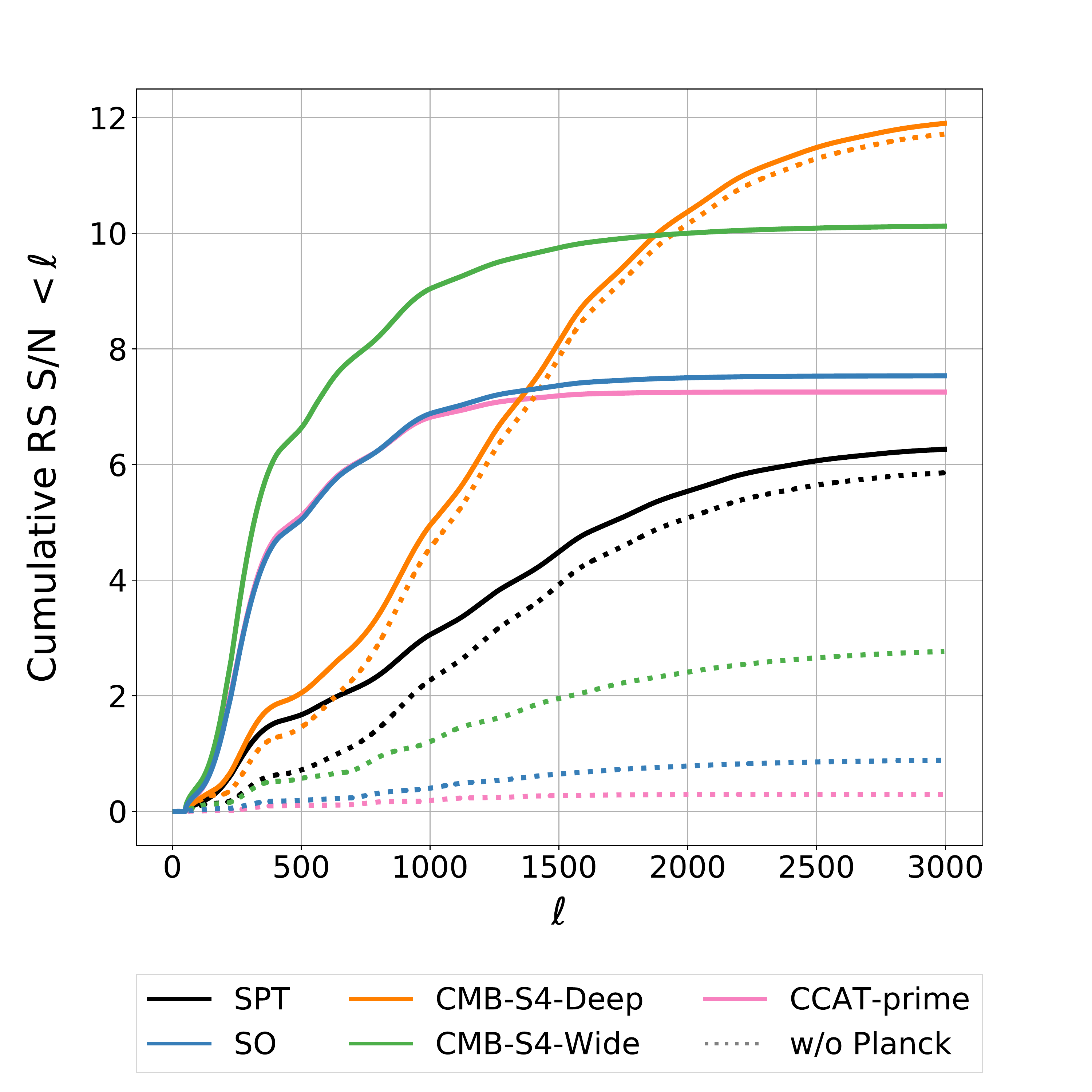}}
\caption{
\label{fig:compare_atm} 
Comparison of the forecasted Rayleigh signal-to-noise for upcoming CMB experiments. \textit{Left:} Achievable signal-to-noise with detector noise as the only component in the noise model. \textit{Right:} Signal-to-noise achievable with both detector noise and atmospheric components included in the noise model. All experiments are assumed to be combined with Planck data. Dotted lines represent the Rayleigh signal-to-noise achievable for each experiment without Planck data. The addition of the atmosphere severely impacts wide experiments, and the majority of their Rayleigh detections come from Planck. The addition of atmosphere also removes low-$\ell$ signal-to-noise from deep experiments, but Planck data only constitutes a small portion of their Rayleigh detections.
}
\end{centering}
\end{figure*}

\begin{figure*}
\begin{centering}
\subfigure[Detector noise and galactic foregrounds]{\label{fig:gal} 
 \includegraphics[width=0.95\columnwidth, trim=0 0 0 65, clip]{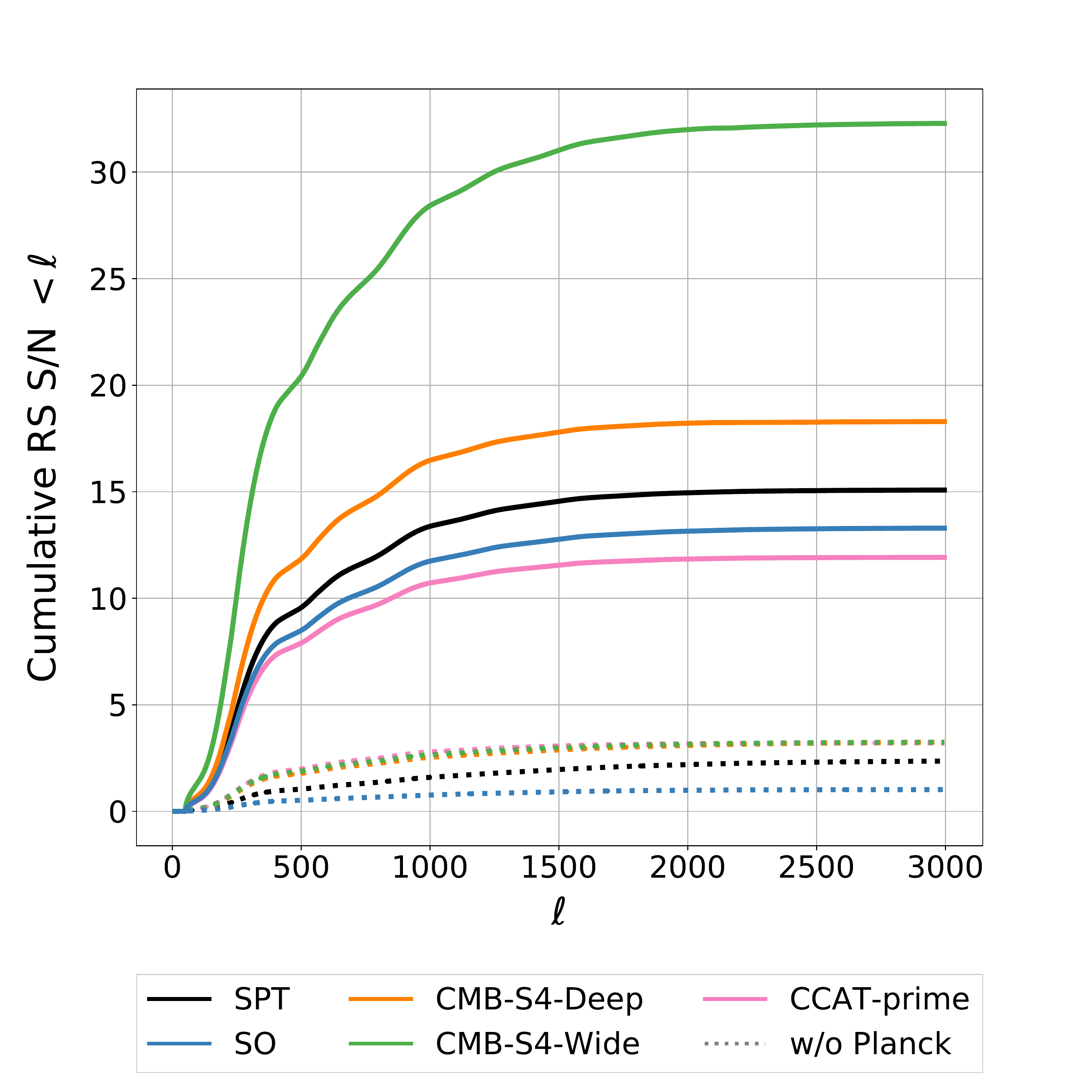}}
\subfigure[Detector noise and extragalactic foregrounds]{\label{fig:exg} 
 \includegraphics[width=0.95\columnwidth, trim=0 0 0 65, clip]{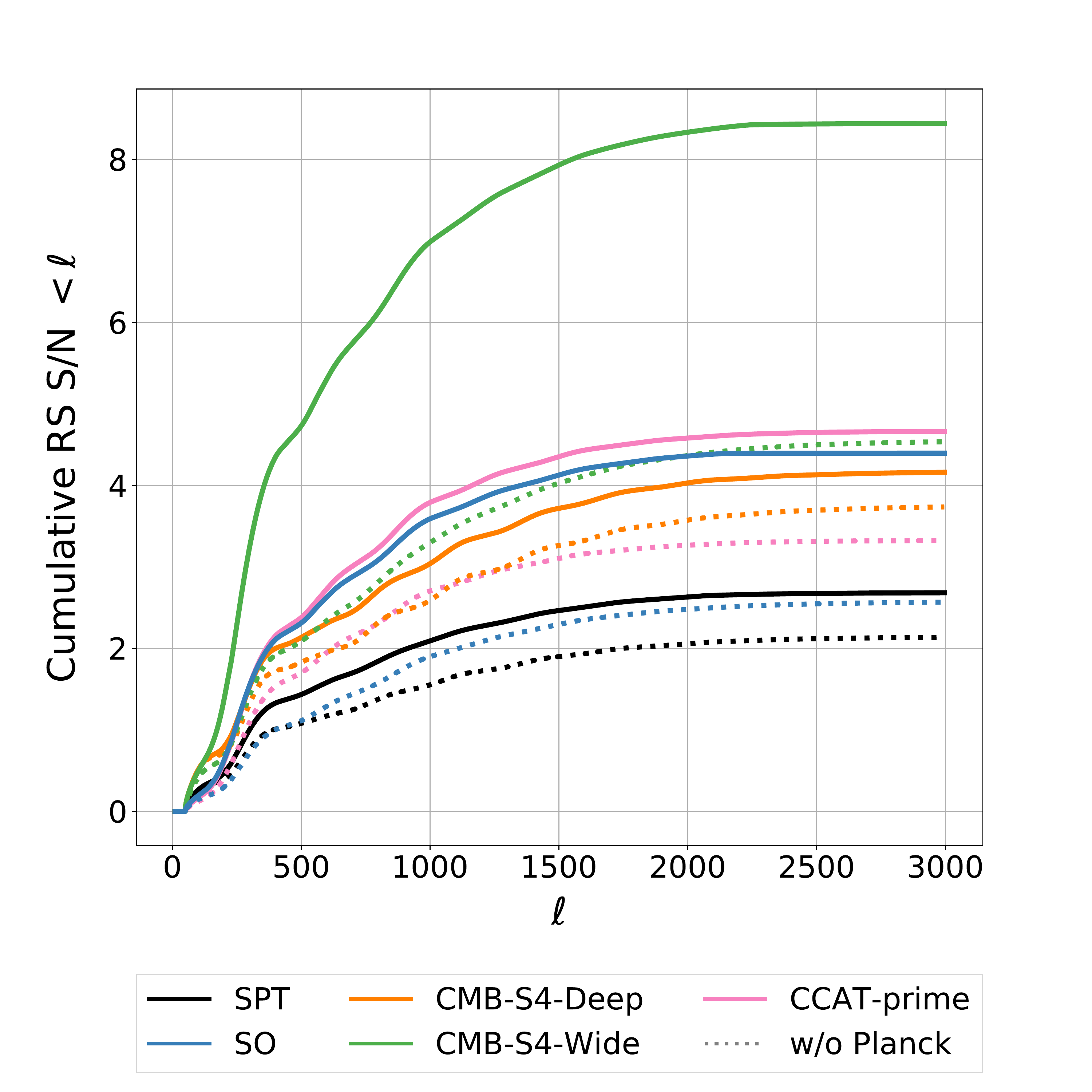}}
\caption{
\label{fig:compare_fg} 
Comparison of the total forecasted Rayleigh signal-to-noise for upcoming CMB experiments with galactic (\textit{Left}) and extragalactic (\textit{Right}) foregrounds included in addition to detector noise. Again all experiments are assumed to be combined with Planck data, and dotted lines represent the achievable Rayleigh signal-to-noise for each experiment without Planck data. The left panel illustrates the ability of Planck data to remove the galactic dust component from CMB maps. CMB-S4-Wide, with its large field, benefits the most from this effect. The right panel reveals that extragalactic foregrounds severely decrease the Rayleigh detection significance of all ground-based experiments, even when Planck data is included. For deep experiments, this loss is more significant than that caused by the atmosphere.
}
\end{centering}
\end{figure*}

The Rayleigh scattering signal-to-noise is dominated by contribution from the primary CMB temperature--Rayleigh temperature cross-spectrum $T \DT$. This is illustrated by Figure \ref{fig:spt4_spectra}, which shows the relative signal-to-noise of each of the four available primary-Rayleigh cross-spectra: $T\DT, T\DE, E\DT,$ and $E\DE$ in relation to the total combined signal-to-noise. The left panel includes no foregrounds, and the right panel includes all foregrounds. As one would expect, multiple pairs of spectra are strongly correlated, meaning that the Rayleigh scattering information contained within each of these signals is not independent. We see this manifest in the total combined Rayleigh signal-to-noise (black line in Figure \ref{fig:spt4_spectra}) being lower than the quadrature sum of the signal-to-noise values of the individual cross-spectra in Figure \ref{fig:spt4_spectra}. These four cross-spectra respond differently to the presence of foregrounds. Largely unpolarized extragalactic foregounds are the limiting noise component for $\langle T \DT \rangle $ and $\langle E \DT \rangle$, which degrade severely between the left and right panels of Figure \ref{fig:spt4_spectra}. The remaining spectra, $\langle T \DE \rangle $ and $\langle E \DE \rangle$, are only midly affected, mostly by the $10\%$ polarized galactic dust component.

\subsection{Forecasts for other upcoming experiments}

\begin{figure*}
    \centering
    \includegraphics[trim = 0 0 0 65, clip,width=1.5\columnwidth]{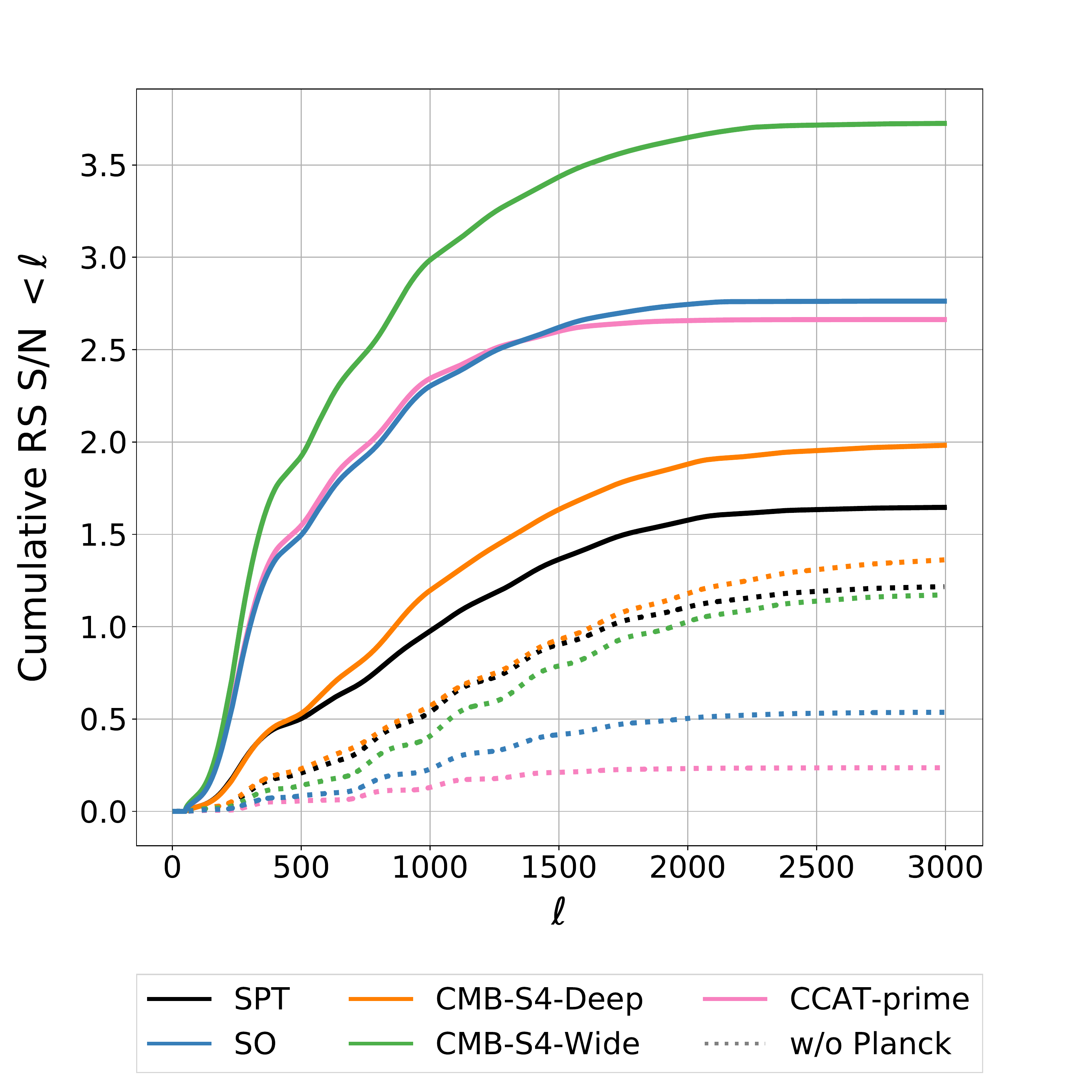}
    \caption{Total forecasted Rayleigh scattering signal-to-noise for upcoming ground-based experiments. As above, dotted lines represent the signal-to-noise for each experiment if Planck data is not included. The inclusion of Planck data majorly benefits wide experiments, which are able to utilize a larger portion of Planck's sky coverage. Without Planck, deep experiments expect slightly more significant Rayleigh scattering detections.} 
    \label{fig:compare_all}
\end{figure*}

\begin{table*}[]
    \centering
    \resizebox{0.95\textwidth}{!}{ \begin{tabular}{|c|c|c|c|c|c|c|c|c|c|c|}
    \hline
          \multirow{2}{*}{Experiment ($f_\text{sky}$)}&  \multicolumn{2}{c|}{Detectors}   &   \multicolumn{2}{c|}{Det. + Atmos.}      & \multicolumn{2}{c|}{Det. + Gal.}      & \multicolumn{2}{c|}{Det. + Exgal.}     &  \multicolumn{2}{c|}{All} \\
        \cline{2-11}
        & w/ Planck & Alone & w/ Planck & Alone & w/ Planck & Alone & w/ Planck & Alone & w/ Planck & Alone\\
        \hline
        SPT (3\%)& 21.2 & 21.1 & 6.3 & 5.9 & 15.1 & 2.4 & 2.7 & 2.1 & 1.6 & 1.2 \\
        \hline
        Simons Obs. (40\%) & 14.1 & 12.0 & 7.5 & 0.9 & 13.3 & 1.0 & 4.4 & 2.6 & 2.8 & 0.5\\
        \hline
        CCAT-prime (44\%) & 12.7 & 10.0 & 7.3 & 0.3 & 11.9 & 3.2 & 4.7 & 3.3 & 2.7 & 0.2\\
        \hline
        CMB-S4 Deep (3\%) & 40.0 & 39.9 & 11.9 & 11.7 & 18.3 & 3.2 & 4.2 & 3.7 & 2.0 & 1.4\\
        \hline
        CMB-S4 Wide (65\%) & 35.9 & 34.6 & 10.1 & 2.8 & 32.3 & 3.2 & 8.4 & 4.5 & 3.7 & 1.2\\
        \hline
        Planck (65\%) & - & 8.7 & - & 8.7 & - & 8.1 & - & 3.7 & - & 3.2\\
        \hline
    \end{tabular}
    }
    \caption{Total forecasted Rayleigh scattering signal-to-noise for upcoming ground-based CMB experiments combined with Planck data. This table summarizes the results displayed in Figures \ref{fig:compare_atm}, \ref{fig:compare_fg}, and \ref{fig:compare_all}. Column labels indicate which noise components are included in the model to produce the forecasts in a given column. For each set of noise components, subcolumns indicate the forecasted Rayleigh scattering signal-to-noise with and without the addition of Planck data. The bottom row shows forecasts for Planck data only, assuming 65\% sky coverage.}
    \label{tab:results_all}
\end{table*}

Figure \ref{fig:compare_atm} shows the Rayleigh signal-to-noise forecasted for SPT-3G+ (black) along with Simons Observatory \cite{simons19} (blue), CCAT-prime \cite{ccat21, choi20} (pink), and the CMB-S4 \cite{cmbs422} wide field survey (green) and deep field survey (orange) in the presence of only detector and atmospheric noise. Detector noise and atmospheric parameters used for each of these experiments are given in Table \ref{tab:instruments}, while galactic foreground estimates are shown in Table~\ref{tab:foregrounds}. All cumulative signal-to-noise values reported in this section for each experiment with and without including Planck data are recorded in Table \ref{tab:results_all}.

Solid colored lines in Figure~\ref{fig:compare_atm} indicate the total Rayleigh signal-to-noise achievable for each experiment when combined with Planck data. In this analysis, each ground-based experiment is combined with the Planck data that overlaps each experiment's observing area on the sky. Thus, wide experiments are able to utilize a larger portion of the available Planck information than are deep experiments. The dotted lines in Figure \ref{fig:compare_atm} indicate the Rayleigh signal-to-noise achievable by each experiment without including Planck data. The left panel of Figure \ref{fig:compare_atm} shows Rayleigh forecasts in the absence of any foregrounds or atmosphere. In this limit, all experiments show significant improvements over Planck in Rayleigh sensitivity. 

The right panel of Figure \ref{fig:compare_atm} shows the effect of adding the atmospheric noise component described in the previous section. The atmospheric noise decreases the detection significance of the wide survey experiments more significantly, such that the resulting Rayleigh detection of a wide experiment comes mostly from the Planck data with which it is combined. Deep experiments lose significant low-ell signal-to-noise, but the majority of the Rayleigh scattering detection for each deep experiment still comes from the experiment itself (not Planck).

The left panel of Figure \ref{fig:compare_fg} shows the Rayleigh forecasts for upcoming ground-based experiments in the presence of detector noise and galactic foregrounds only. Galactic foregrounds do not affect the achievable Rayleigh signal-to-noise of ground-based experiments as much as the atmospheric contamination. With galactic dust and synchrotron emission as the only foregrounds, all upcoming ground-based CMB experiments perform relatively similarly when combined with Planck, with the exception of CMB-S4-Wide, which performs significantly better. Without Planck, the Rayleigh detection significance of all ground-based experiments falls to a similar 3-4-$ \sigma$ level, highlighting the ability of Planck data to remove galactic dust contamination during component separation. CMB-S4-Wide, with the largest observing field among the experiments considered here, benefits the most from combination with Planck data.

The right panel of Figure \ref{fig:compare_fg} shows the Rayleigh forecasts in the presence of detector noise and extragalactic foregrounds (tSZ, CIB, and extragalactic radio sources) only. This panel illustrates the significant impact of extragalactic foregrounds on Rayleigh scattering detections, even when Planck data is utilized. For wide experiments combined with Planck, the loss in detection significance due to extragalactic foregrounds alone is approximately equal to the loss due to atmosphere. For deep experiments combined with Planck, this loss is significantly more severe than atmospheric loss. Comparing to the left panel, it is clear that Planck is not nearly as successful at removing extragalactic foregrounds during component separation as it is at removing galactic foregrounds.

Including all of the above noise components in our model, we produced total forecasts for Rayleigh scattering signal-to-noise for upcoming experiments in the presence of atmospheric, galactic, and extragalactic foregrounds. These total forecasts are presented in Figure \ref{fig:compare_all}. All forecasted signal-to-noise values are shown in Table \ref{tab:results_all}. These forecasts indicate that, in combination with Planck data, all upcoming ground-based CMB experiments can expect a Rayleigh scattering detection with a signal-to-noise of roughly 1-4. For wide experiments, the majority of this detection comes from Planck data, as indicated by the dotted lines. Though deep experiments can expect slightly lower signal-to-noise than wide experiments, their Rayleigh scattering detections come mostly from the experiments themselves. Without Planck, the highest-significance Rayleigh scattering detections of 1.5-2 come from deep experiments. It is also relevant to note that this model predicts that a roughly 3-$\sigma$ Rayleigh scattering detection is potentially present in the Planck dataset corresponding to the CMB-S4-Wide observing patch, which encompasses 65\% of the sky. This is backed up by the forecasted signal-to-noise values for Planck alone with $f_\text{sky}=0.65$, which are shown in the last row of Table \ref{tab:results_all}. 

Of the components present in our extragalactic foregrounds model, we found the CIB to be the largest limiter of total achievable Rayleigh signal-to-noise.

\begin{figure*}
\begin{centering}
\subfigure[Detector noise and atmosphere only (no Planck)]{\label{fig:atmos_corr_det_np} 
 \includegraphics[width=0.95\columnwidth, trim=0 0 0 65, clip]{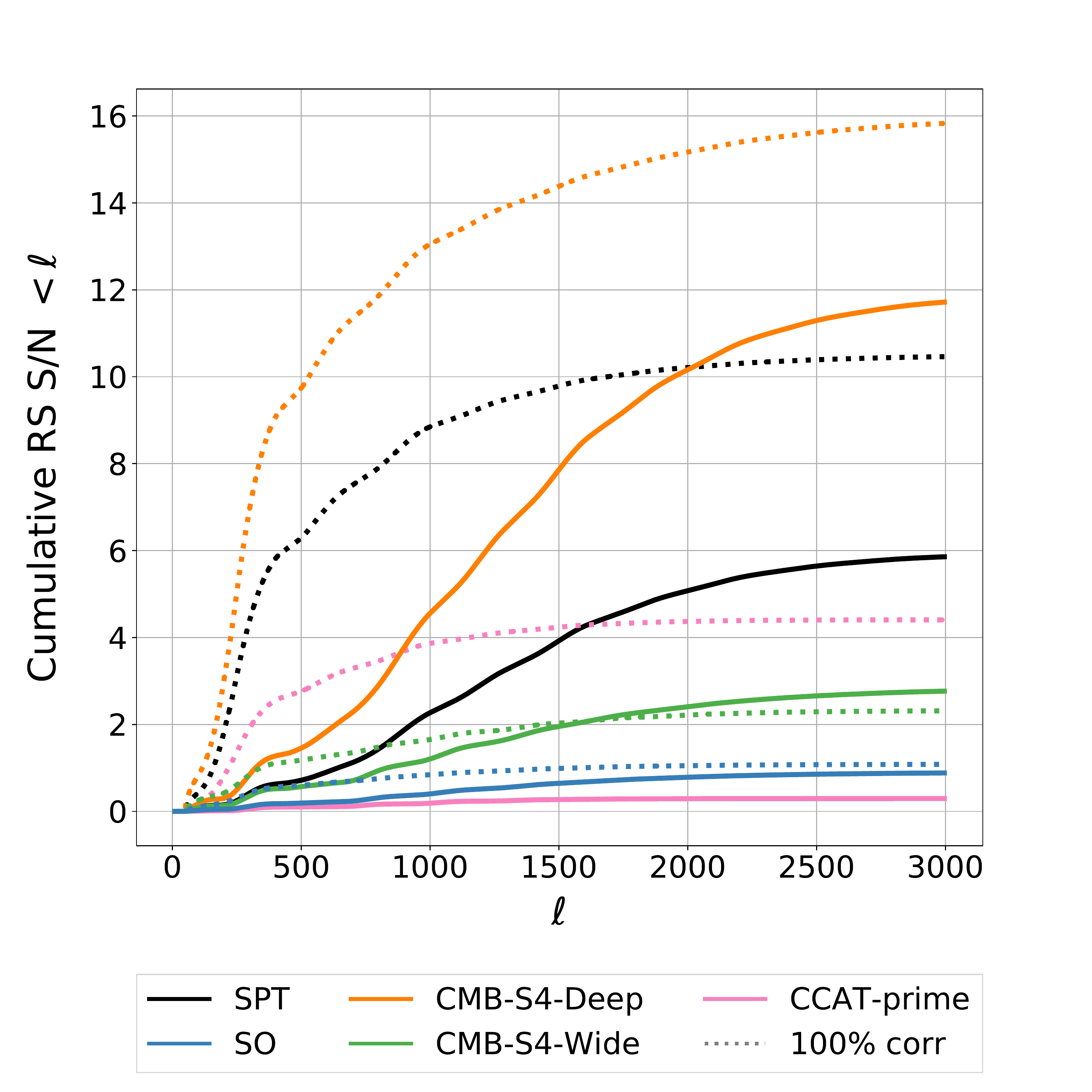}}
\subfigure[All foregrounds combined with Planck]{\label{fig:atmos_corr_all_planck} 
 \includegraphics[width=0.95\columnwidth, trim=0 0 0 65, clip]{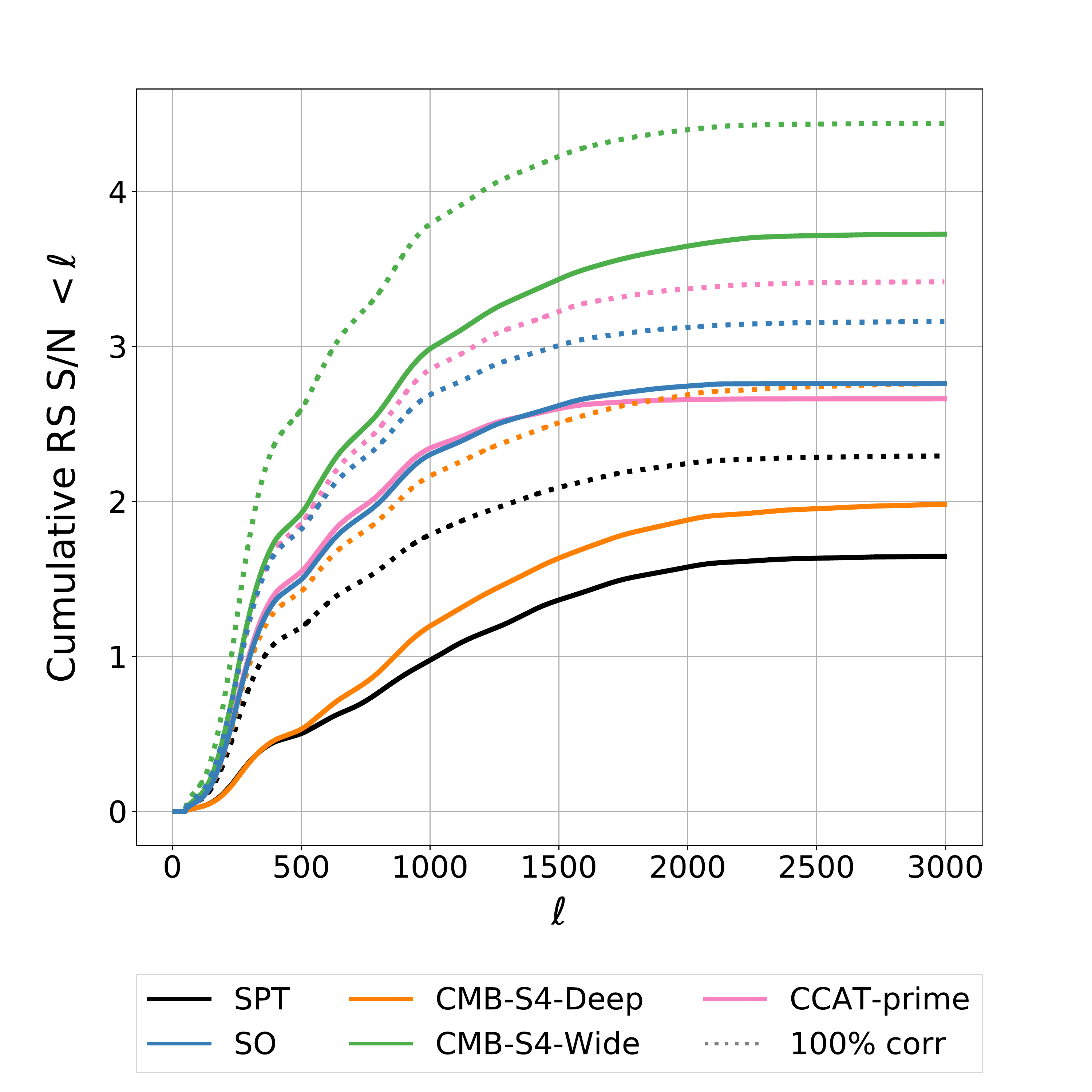}}
\caption{
\label{fig:atmos_corr} 
Rayleigh forecasts when the atmosphere is taken to be totally correlated between bands. \textit{Left:} Forecasts including only detectors and correlated atmosphere. In this plot, experiments are not combined with Planck so that the effect of correlated atmosphere may be clearly seen. Solid lines correspond with the dotted lines in the right panel of Figure \ref{fig:compare_atm}, and dotted lines represent the same forecasts with a fully correlated atmosphere. \textit{Right:} Forecasts including Planck data, all foregrounds and a fully correlated atmosphere. Solid lines here correspond to the solid lines in Figure \ref{fig:compare_all}, and dotted lines represent the same forecasts with a fully correlated atmosphere.
}\label{fig:atmos_corr}
\end{centering}
\end{figure*}

\subsection{Atmospheric correlation}
\label{sec:atmocorr}
As noted in Section~\ref{sec:atmo}, in our fiducial forecasting pipeline we assume low-$\ell$ noise from the atmosphere to be uncorrelated between frequency bands. Depending on the specific experiment configuration, and in the limit that the low-$\ell$ noise from the atmosphere comes entirely from clouds of water vapor that are optically thin at all observing frequencies, this contribution could in principle be nearly 100\% correlated between detectors and frequency bands. One promising path towards mitigating atmospheric contamination recalls early CMB/tSZ experiments such as SuZIE \cite{holzapfel97a}, in which the atmosphere is at least partially mitigated by forming linear combinations of channels that are least sensitive to atmosphere---i.e., treating the atmosphere in the same way we treat correlated foregrounds in this work \cite{mauskopf00}. 

We produce an alternate set of forecasts in which the atmospheric contribution is 100\% correlated between bands. The most straightforward way to achieve this would be to modify Equation~\ref{eqn:atmocov} to read
\begin{eqnarray}
\bm{N}^\text{atmos}_\ell (\nu_i,\nu_j) &=& \sqrt{N^\text{det}_i N^\text{det}_j} \times \\
\nonumber && \Big( \frac{\ell_\text{knee}(\nu_i)  }{\ell} \Big)^{\alpha_i/2}
\Big( \frac{\ell_\text{knee}(\nu_j)  }{\ell} \Big)^{\alpha_j/2}.
\label{eqn:atmocovcorr}
\end{eqnarray}
The problem with this formulation is that the values of $\ell_\text{knee}$ (in temperature) for the various upcoming experiments were estimated assuming that the atmospheric noise will integrate down at least partially as the number of detectors is increased. If atmospheric noise is instead 100\% correlated across all detectors and bands, its power spectrum in a given band will be independent of detector number. To create a self-consistent atmospheric noise covariance matrix for the fully correlated case, we must scale the amplitude back up by the amount it was assumed to scale down in the uncorrelated case. 

The values of $\ell_\text{knee}$ for the future South Pole experiments SPT-3G+ and CMB-S4 Deep are taken directly from measurements in SPT-3G; as such, they implicitly assume that the atmospheric noise will integrate down with the number of detectors. The values of $\ell_\text{knee}$ for the future Chile experiments SO, CCAT-prime, and CMB-S4 Wide are calculated using the SO Noise Calculator (as described in \cite{simons19}), which starts with noise power spectra measured with ACTPol and assumes that independent camera sub-modules or optics tubes will see independent atmosphere---i.e., that the atmospheric noise will scale from ACTPol to a future instrument by the inverse number of optics tubes. There is also a factor-of-2 reduction assumed from the larger focal planes of the future instruments. Our self-consistent model for atmospheric noise covariance in the fully correlated case thus looks like
\begin{eqnarray}
\bm{N}^\text{atmos}_\ell (\nu_i,\nu_j) &=& \sqrt{f_i N^\text{det}_i f_j N^\text{det}_j} \times \\
\nonumber && \Big( \frac{\ell_\text{knee}(\nu_i)  }{\ell} \Big)^{\alpha_i/2}
\Big( \frac{\ell_\text{knee}(\nu_j)  }{\ell} \Big)^{\alpha_j/2},
\label{eqn:atmocovcorr2}
\end{eqnarray}
where $f_i$ is a scaling factor that is equal to $n_i^\text{det}/n_i^\text{det, SPT-3G}$ (where $n_i^\text{det}$ is the number of detectors in band $i$) for the future experiments at the South Pole and $2 \times n_i^\text{tube}$ (where $n_i^\text{tube}$ is the number of optics tubes in band $i$) for the future experiments in Chile. Finally, we note that because SPT-3G and SPT-3G+ will not observe simultaneously, we zero the atmospheric noise correlations between the SPT-3G and SPT-3G+ bands in the SPT covariance matrix.

The dotted lines in Figure \ref{fig:atmos_corr} show the effects of the fully correlated atmosphere model relative to the fully uncorrelated model used above. These represent two extremes of atmosphere correlation, meaning that with Planck data and all foregrounds included, the true Rayleigh scattering signal-to-noise should lie somewhere between the solid and dotted lines in the right-hand panel of Figure \ref{fig:atmos_corr}.

\section{Conclusions}

 A ground-based Rayleigh scattering detection is challenging in that it requires an experiment to have high sensitivity at frequencies beyond the peak of the CMB blackbody spectrum as well as the ability to mitigate both atmospheric and astrophysical foreground contamination. With many upcoming CMB ground-based experiments proposing low-noise, high-frequency cameras, a first detection of Rayleigh scattering is moving closer into reach. Our Rayleigh scattering forecasting pipeline, based on the constrained linear combination method described in \cite{remazeilles18}, indicates that, though upcoming experiments will be severely limited by atmospheric emission and extragalactic foregrounds, a first Rayleigh scattering detection may still be possible in the upcoming decade if experiments combine their data with Planck and place high priority on understanding and removing both atmospheric contamination and that from extragalactic foregrounds. Extragalactic foregrounds, particularly the CIB, strongly limit the achievable significance of a Rayleigh scattering detection. This effect is approximately equal to that of the atmosphere for wide experiments, and exceeds the effect of the atmosphere for deep experiments. Thus it is vital for future ground-based Rayleigh scattering detections that attention be paid to understanding and mitigating extragalactic foreground contamination.



When all noise components are included in our model, significant Rayleigh scattering detections are only achievable if ground-based experiments combine their data with the Planck data that overlaps their observation patch. This is particularly true for wide experiments, for whom the majority of the Rayleigh scattering detection comes from the Planck data overlapping their large observation fields. Table \ref{tab:results_all} summarizes the forecasted Rayleigh signal-to-noises for each experiment combined with Planck, where quantities in parentheses indicate how much of each detection comes from Planck data. Without the addition of Planck data, we have found that upcoming experiments can expect a Rayleigh scattering detection signficance of around 1-$\sigma$. This is in agreement with the Rayleigh forecasts presented for CCAT-prime alone in \cite{ccat21}, and in fact, our Rayleigh signal-to-noise forecast for CCAT-prime without Planck with all foregrounds included ($S/N \approx 0.3$) matches that presented in \cite{ccat21}. Our forecasts for Simons Observatory, CCAT-prime, and CMB-S4-Wide without Planck and with only atmosphere included also roughly match those presented in \cite{beringue21}. For deep experiments, though the forecasted Rayleigh detection significance when including Planck data is lower than that of wide experiments, the majority of the signal-to-noise in these deep experiment detections comes from the deep experiments themselves (Figure \ref{fig:compare_all}). 

We find that Planck data alone, with $f_\text{sky}=0.65$, may already contain a roughly 3-$\sigma$ Rayleigh scattering detection, as shown in the last row of Table \ref{tab:results_all}. A higher Rayleigh detection may also be achievable by combining deep ground-based experiments with all available Planck data, rather than just the Planck data that overlaps these experiments' $\fsky = 0.03$ observing patches. 
Further progress on atmospheric and CIB removal, beyond that considered here, will likely be necessary for current and planned ground-based experiments to significantly increase the Rayleigh scattering detection significance beyond what should be achievable from Planck data alone.

\begin{acknowledgements}
KD is supported by the Graduate Instrumentation Research Award through the Department of Energy, Office of High Energy Physics. The South Pole Telescope program is supported by the National Science Foundation (NSF) through the grant OPP-1852617. Partial support is also provided by the Kavli Institute of Cosmological Physics at the University of Chicago. Partial support for SPT-3G+ development is provided by NSF grant OPP-2117894. 
Work at Fermi National Accelerator Laboratory, a DOE-OS, HEP User Facility managed by the Fermi Research Alliance, LLC, was supported under Contract No. DE-AC02-07CH11359. 
Argonne National Laboratory's work was supported by the U.S. Department of Energy, Office of High Energy Physics, under contract DE-AC02-06CH11357. ZP is supported by Argonne National Laboratory under award LDRD-2021-0186. 
W.L.K.W is supported in part by the Department of Energy, Laboratory Directed Research and Development program and as part of the Panofsky Fellowship program at SLAC National Accelerator Laboratory, under contract DE-AC02-76SF00515.
The Melbourne group acknowledges support from the Australian Research Council's Discovery Projects scheme (DP210102386). 
During the course of this work, we became aware of concurrent work by Zhu et al. (\textit{in preparation}). We thank the authors of this paper for useful conversations.
\end{acknowledgements}


\begin{table*}[]
    \centering
    \resizebox{0.8\textwidth}{!}{  \begin{tabular}{|c|c|c|c|c|c|c|c|c|}
        \hline
        Instrument &  Band & Beam & $T$ Map Depth & $E$ Map Depth & $\ell_{\text{knee},T}$ & $\alpha_T$ & $\ell_{\text{knee},E}$ & $\alpha_E$ \\
                & (GHz) & (arcmin)& ($\mu$K-arcmin) & ($\mu$K-arcmin) & & & & \\
        \hline
        SPT-3G          &  95    & 1.7   &  2.7   & 3.8     & 1200     & -4.2     & \multirow{3}{*}{200}     & -2.6\\
        ($\fsky=0.03$)& 150    & 1.2   &  2.2   & 3.1     & 1900     & -4.1     &   & -2.2\\
                        & 220    & 1.1   &  8.8   & 12.4     & 2100    & -3.9     & & -2.2\\
        \hline
        SPT-3G+         & 225    & 0.8   &  2.9   & 4.1    & \multirow{2}{*}{2100}      & \multirow{3}{*}{-3.9}     & \multirow{3}{*}{200}     & \multirow{3}{*}{-2.2}\\
        ($\fsky=0.03$)& 285    & 0.6   &  5.6   & 7.9    &  & &  & \\
        \cline{6-6}
                        & 345    & 0.5   & 28     & 39.6    & 2600     & &  & \\
        \hline
        Simons Obs.     &  27    & 7.4   & 52     & 74      & \multirow{2}{*}{400}   & \multirow{6}{*}{-3.5} & \multirow{6}{*}{700}     & \multirow{6}{*}{-1.4}\\
         ($\fsky=0.40$)& 39    & 5.1   & 27     & 38      & & &  & \\
         \cline{6-6}
                        &  93    & 2.2   &  5.8   & 8.2    & 1900      & &  & \\
                        & 145    & 1.4   &  6.5   & 9.2    & 3900   & &  & \\
                        & 225    & 1.0   & 15     & 21.2    & 6700     & & & \\
                        & 280    & 0.9   & 37     & 52.3    & 6800     & & & \\
        \hline
        CCAT-p          & 220    & 1.0   & 15     & 21.2    & 7300     & \multirow{5}{*}{-3.5}  & \multirow{5}{*}{700}     & \multirow{5}{*}{-1.4}\\
        ($\fsky=0.50$)& 280    & 0.8   & 28     & 39.6    & 8800     & & & \\
                        & 350    & 0.6   &107     & 151    & 10600     & & & \\
                        & 410    & 0.5   &407     & 576    & 8200      & & & \\
                        & 850    & 0.3   &$6.8 \times 10^5$ & $9.6 \times 10^5$    & 4600 & & & \\
        \hline
         CMB-S4-Wide   & 27    & 7.4   & 21.5     & 30.4   & \multirow{2}{*}{400}      & \multirow{6}{*}{-3.5} & \multirow{6}{*}{700}     & \multirow{6}{*}{-1.4}\\
         (Chilean LAT)  & 39    & 5.1   & 11.9     & 16.8   & & & & \\
         \cline{6-6}
                      &  93    & 2.2   & 1.9    & 2.7    & 1900      & & & \\
                       & 145    & 1.2   & 2.1    & 2.9    & 3900      & & & \\
        ($\fsky=0.65$)& 225    & 0.9   & 6.9    & 9.7    & 6700      & & & \\
                        & 278    & 0.7   & 17     & 23.8   & 6800      & & & \\
        \hline
        CMB-S4-Deep     & 20    & 11.4   & 8.7    & 12.3    & \multirow{3}{*}{400}    & \multirow{4}{*}{-4.2}     & \multirow{4}{*}{150}     & \multirow{3}{*}{-2.7}\\
        (S. Pole TMA)   & 27    & 8.4   & 5.1   & 7.1    & & & & \\
        ($\fsky=0.03$)& 39    & 5.8   & 3.3    & 4.6   & & & & \\
        \cline{6-6}\cline{9-9}
                          &  95    & 2.5   & 0.5    & 0.71    & 1200     & &     & -2.6\\
                          \cline{7-9}
                         & 150    & 1.6   & 0.5    & 0.66    & 1900     & -4.1     & \multirow{3}{*}{200}     & \multirow{3}{*}{-2.2}\\
                         \cline{6-6}
                        & 220    & 1.1   & 1.5    & 2.05    & \multirow{2}{*}{2100} & -3.9     & & \\
                        & 285    & 1.0   & 3.4    & 4.85    &  & -3.9     & & \\
        \hline
        
    \end{tabular}
    }
    \caption{Instrument and atmospheric parameters for all ground-based experiments considered in this analysis. $N_\text{det}$ in Section \ref{noise} is defined as the square of the map depth. SPT-3G map depth values come from \cite{bender18}, while SPT-3G+ values are calculated using design detector noise and assuming four years of observation with similar efficiency to SPT-3G. SPT-3G atmospheric noise parameters come from on-sky measurements, and SPT-3G+ values are scaled from these using PWV values in each band. CCAT-prime values come from \cite{ccat21}. CMB-S4 Deep map depths, bands, and beams are taken from the CMB-S4 wiki.\footnote{\url{https://cmb-s4.uchicago.edu/wiki/index.php/Delensing_sensitivity_-_preliminary_results\#V3R0}} Since atmospheric parameters on that wiki were not updated at the time of writing, CMB-S4 Deep atmospheric parameters were assumed to be identical to those of the corresponding SPT bands. Official CMB-S4 Wide values were also not publicly available at the time of writing, so values were taken from Table VIII of \cite{beringue21}. Simons Observatory map depths, bands, and beams come from \cite{simons19}, and atmospheric parameters were assumed to be identical to those of the corresponding CMB-S4 Wide bands. Since CMB-S4 Wide and CCAT-prime use a different atmospheric model to the one described in Section \ref{noise}, parameters have been converted to ones that produce equivalent atmospheric noise in our model. Planck values for map depth, though not included in Table \ref{tab:instruments}, are taken from Table 4 of \cite{planck20b}. }
    \label{tab:instruments}
\end{table*}


\begin{table*}[]
    \centering
    \resizebox{0.7\textwidth}{!}{  
    \begin{tabular}{|c |c|c|c|c|c|}
    \hline
    Foreground & Parameter & Deep TT & Wide TT & Deep EE & Wide EE \\
    \hline
    
    \multirow{4}{*}{Galactic} & $A_{\text{dust},145}$ [$\mu$K$^{2}$] & 3.253&  1168 & 0.048& 1.161 \\
    \cline{2-6}
    & $\alpha_{\text{dust},145}$ & -0.400 & -0.246 & -0.400 & -0.371\\
    
    \cline{2-6}
     & $A_{\text{synch},93}$ [$\mu$K$^{2}$] & 0.005 & 0.055 & 0.001 & 0.010\\
     \cline{2-6}
    & $\alpha_{\text{synch},93}$ & \multicolumn{4}{c|}{-0.4} \\
    
    \hline
      \multirow{12}{*}{Extragalactic} & $A_{\text{tSZ},150}$ [$\mu$K$^{2}$] & \multicolumn{2}{c|}{4} & \multicolumn{2}{c|}{\multirow{12}{*}{N/A}}\\
      \cline{2-4}
     & $\alpha_{\text{tSZ},150}$ & \multicolumn{2}{c|}{0} & \multicolumn{2}{c|}{} \\
     \cline{2-4}
     
      & $A_{\text{Cl-lo}z,220}$ [$\mu$K$^{2}$] & \multicolumn{2}{c|}{40}& \multicolumn{2}{c|}{}\\
      \cline{2-4}
      & $\alpha_{\text{Cl-lo}z,220}$ & \multicolumn{2}{c|}{0.8} & \multicolumn{2}{c|}{}\\
     \cline{2-4}
     
      & $A_{\text{Cl-hi}z,220}$ [$\mu$K$^{2}$] & \multicolumn{2}{c|}{20} & \multicolumn{2}{c|}{} \\
      \cline{2-4}
      & $\alpha_{\text{Cl-hi}z,220}$ & \multicolumn{2}{c|}{0.8} & \multicolumn{2}{c|}{}\\
     \cline{2-4}
     
      & $A_{\text{Po-lo}z,220}$ [$\mu$K$^{2}$] & \multicolumn{2}{c|}{20} & \multicolumn{2}{c|}{}\\
      \cline{2-4}
      & $\alpha_{\text{Po-lo}z,220}$ & \multicolumn{2}{c|}{2} & \multicolumn{2}{c|}{}\\
     \cline{2-4}
     
      & $A_{\text{Po-hi}z,220}$ [$\mu$K$^{2}$] & \multicolumn{2}{c|}{50} & \multicolumn{2}{c|}{}\\
      \cline{2-4}
      & $\alpha_{\text{Po-hi}z,220}$ & \multicolumn{2}{c|}{2} & \multicolumn{2}{c|}{}\\
     \cline{2-6}
     
      & $A_{\text{Radio},150}$ [$\mu$K$^{2}$] & \multicolumn{2}{c|}{0.17} & \multicolumn{2}{c|}{$8 \times 10^{-5}$}\\
      \cline{2-6}
      & $\alpha_{\text{Radio},150}$ & \multicolumn{4}{c|}{2}\\
     \hline
    \end{tabular}
    }
    \caption{Anchor values for galactic and extragalactic foreground amplitudes and power law slopes. Galactic foregrounds include galactic dust and galactic synchrotron radiation, while extragalactic foregrounds include the thermal Sunyaev Zel{'}dovich effect, the four-component CIB model presented in Section IV.D, and extragalactic radio sources. The models for each of these components are detailed in \S~\ref{sec_gal_sources} and \S~\ref{sec_extragal_sources}. These values are scaled to other frequency bands as described in Appendix ~\ref{appendix_gal_foregrounds_scaling} and \ref{appendix_extragal_foregrounds_scaling}.}
    \label{tab:foregrounds}
\end{table*}

\appendix 

\section{Noise model parameters}
\label{sec_noise_params}

\subsection{Instrument and atmospheric parameters}
\label{sec_ins_atmos_noise_params}

Table \ref{tab:instruments} lists the values of beam size (FWHM), map depth, and atmospheric model parameters ($\ell_\text{knee}$ and $\alpha$) for each band of each ground-based instrument considered in this analysis. Assumed full-survey map depths for SPT-3G are taken from \cite{bender18}; map depths for SPT-3G+ are calculated from design detector noise values and assuming four years of observation with efficiency similar to SPT-3G. Atmospheric parameter values for SPT-3G are estimated from on-sky data; values for SPT-3G+ at 220~GHz are assumed to be identical to SPT-3G, while numbers for higher-frequency bands are scaled using the measured levels of precipitable water vapor (PWV) at the South Pole integrated over the design SPT-3G+ bands. 

Sources for other experiments' values are given in the caption to Table \ref{tab:instruments}. Map depth values for Planck, though not included in Table \ref{tab:instruments}, are taken from Table 4 of \cite{planck20b}. Some experiments use a different atmospheric model in which $\ell_\text{knee}$ is fixed and $N_\text{atmos} = N_\text{red} (\frac{\ell_\text{knee}}{\ell})^\alpha + N_\text{white}$, where $N_\text{white}$ is the detector noise. For these experiments, we convert their parameters to the equivalent ones in our model, using:

\begin{equation}
    \ell_\text{knee} = \ell_\text{fixed} \left(\frac{N_\text{red}}{N_\text{white}} \right )^\frac{1}{\alpha},
\end{equation}
where $\ell_\text{fixed}$ is the fixed value of $\ell_\text{knee}$ used in their atmospheric model.

\subsection{Galactic dust and synchrotron amplitudes}  
\label{appendix_gal_foregrounds_scaling}
We rely on the publicly available 
Python Sky Model (pySM) simulations \citep{thorne17, zonca21} developed based on the \planck{} Sky Model code \citep{delabrouille13} for galactic foregrounds. 
The approach is similar to the one followed in \citep{raghunathan22}.
For both the Deep and Wide surveys, we estimate the power spectrum of the galactic dust and synchrotron signals in pySM, both in temperature $C_{\ell}^{TT}$ and polarization $C_{\ell}^{EE}$. 
Since $C_{\ell}^{TE}=0$ in pySM, we set the TE correlation using the geometric mean of the two signals as $C_{\ell}^{TE} = \rho_{TE}^{\rm gal} \sqrt{C_{\ell}^{TT} C_{\ell}^{EE}}$ with $ \rho_{TE}^{\rm gal} = 0.35$ for all galactic foregrounds \citep{planck18-11}. We use the pySM \texttt{S0\_d0} dust and \texttt{S0\_s0} synchrotron models in this work. 

We fit a power law of the form $D_{\ell} = A \left( \dfrac{\ell}{80} \right)^{\alpha}$ to determine the dust and synchrotron amplitudes at our reference frequencies of 145 and 93 GHz, respectively, and we scale those amplitudes to other bands as:
\begin{equation}
\label{eq_gal_dust}
C_{\ell, \nu_1 \nu_2} = C_{\ell, \nu_{0} \nu_{0}}\ \epsilon_{\nu_{1}, \nu_{2}}\ \dfrac{\eta_{\nu_{1}} \eta_{\nu_{2}}}{\eta_{\nu_{0}} \eta_{\nu_{0}}}, 
\end{equation}
where $\nu_{0}=$ 145 or 93 GHz, and $\nu_1$, $\nu_2$ correspond to frequency bands listed in Table~\ref{tab:instruments}. The terms $\epsilon_{\nu_1}$ and $\epsilon_{\nu_2}$ in Eq.(\ref{eq_gal_dust}) encode the conversion of radiance to equivalent fluctuation temperature of a 2.7K blackbody: 
\begin{eqnarray}
\label{eq_dust_epsilon}
\epsilon_{\nu_{1}, \nu_{2}} =  \left. \frac{\dfrac{d B_{\nu_{0}}}{dT} \dfrac{d B_{\nu_{0}}}{dT}} {\dfrac{d B_{\nu_{1}}}{dT}\dfrac{d B_{\nu_{2}}}{dT}} \right|_{T = T_{\rm CMB}},
\end{eqnarray}
while $\eta_{\nu}$ represents the spectral energy distribution of either dust or synchrotron. For dust we use a modified blackbody of the form
\begin{eqnarray}
\label{eq_dust_freq_dep}
\eta_\nu = \nu^{\beta_{d}}\ B_{\nu}(T_{d}),
\end{eqnarray}
with $\beta_{d} = 1.6$, and $T_{d}$ = 19.6 K, while for synchrotron we assume a power law in frequency 
\begin{equation}
\eta_\nu = \nu^{2+\beta_s}
\label{eqn:synchscale}
\end{equation}
with $\beta_s = -3.10$.

For simplicity, we assume in the estimation of dust and synchrotron amplitudes that the observing regions for the three ``wide'' experiments (SO, CCAT-prime, and CMB-S4 Wide) are identical and equal to the 57\% of sky available between decl.~68$^\circ$ and 25$^\circ$ and with galactic latitude $b > 10^\circ$. This means that the assumed galactic foreground amplitudes will be slightly pessimistic for SO and CCAT-prime and slightly optimistic for CMB-S4 Wide. Similarly, we assume the observing regions for the two ``deep" experiments (SPT-3G/3G+ and CMB-S4 Deep) are identical and equal to the SPT-3G region defined by $-50^\circ < \mathrm{R.A.} < 50^\circ$ and $-70^\circ < \mathrm{decl.} < -42^\circ$. The galactic foreground parameters derived from this procedure are given in Table \ref{tab:foregrounds}.

\subsection{Extragalactic foreground amplitudes}
\label{appendix_extragal_foregrounds_scaling}
Like the galactic dust model, both the tSZ and CIB models are defined at fiducial frequencies, and the amplitudes are scaled to other frequency bands. The CIB clustered and Poisson amplitudes are defined in Section III.D at 220 GHz and are scaled in exactly the same way as the galactic dust amplitudes above, but with
\begin{equation}
\eta_\nu = B \left ( \nu, \frac{T_\text{CIB}}{1+z} \right ) \nu^2 ,
\end{equation}
where $T_\text{CIB} = 30$K, and $z$ is one of $z_\text{low} = 0.5$ or $z_\text{hi} = 3.5$ corresponding to the low and high redshift CIB components respectively. The clustered and Poisson CIB components are considered unpolarized.

The tSZ amplitude is defined in Section III.D at 150 GHz and is scaled to other bands using the tSZ spectral shape relative to $dB(\nu)/dT_\text{CMB}$:
\begin{equation}
f(\nu)_\text{tSZ} = x \frac{e^x + 1}{e^x - 1} - 4,
\end{equation}
where $x = \nu / 56.8$ GHz. The tSZ amplitude at frequency $\nu$ is:
\begin{equation}
A(\nu)_{\text{tSZ}} = \frac{A_{0, \text{tSZ}}}{f(\nu)_{SZ}^2}.
\end{equation}
We assume the tSZ is unpolarized, and thus all tSZ amplitudes for polarization are zero.
Finally, the extragalactic radio source amplitude, also defined in Section III.D scales to other frequency bands in a similar way to Equation \ref{eqn:synchscale}, but with the $2+\beta_s$ exponent replaced by $-0.7$ (e.g., \cite{mocanu13}). Extragalactic radio sources are considered $3\%$ polarized following \cite{datta18}, \cite{gupta19}, so that extragalactic radio amplitudes for E polarization are $4.5 \times 10^{-4}$ times the extragalactic radio amplitudes for temperature.

\bibliographystyle{unsrtnat}
\bibliography{main.bib}

\end{document}